\documentclass[aps,prb,twocolumn,,groupedaddress,notitlepage,showpacs,floatfix,superscriptaddress]{revtex4-1}
\usepackage{times,amsmath,amsfonts,amssymb,mathrsfs,graphics,graphicx,color,comment,bm}
\usepackage[dvips]{epsfig}
\usepackage[pdfstartview=FitH]{hyperref}
\usepackage{natbib}
\usepackage{appendix}
\usepackage{epstopdf}
\hypersetup{
colorlinks=true,       	
linkcolor=blue,          	
citecolor=blue,        
filecolor=blue,      	
urlcolor=blue,           	
runcolor=blue
}
\newcommand{\ignore}[1]{}
\let\oldsqrt\sqrt
\def\sqrt{\mathpalette\DHLhksqrt}
\def\DHLhksqrt#1#2{%
\setbox0=\hbox{$#1\oldsqrt{#2\,}$}\dimen0=\ht0
\advance\dimen0-0.2\ht0
\setbox2=\hbox{\vrule height\ht0 depth -\dimen0}%
{\box0\lower0.4pt\box2}}

\DeclareFontFamily{OT1}{pzc}{}
\DeclareFontShape{OT1}{pzc}{m}{it}%
              {<-> s * [1.25] pzcmi7t}{}
\DeclareMathAlphabet{\mathpzc}{OT1}{pzc}%
                                 {m}{it}

\begin{document}
\title{Topological phases of the compass ladder model}

\author{R. Haghshenas}
\affiliation{Department of Physics, Sharif University of Technology, Tehran 14588-89694, Iran}
\email{haghshenas@physics.sharif.edu}

\author{A. Langari}
\affiliation{Department of Physics, Sharif University of Technology, Tehran 14588-89694, Iran}
\affiliation{Center of Excellence in Complex Systems and Condensed Matter, Sharif University of Technology, Tehran 14588-89694, Iran}

\author{A. T. Rezakhani}
\affiliation{Department of Physics, Sharif University of Technology, Tehran 14588-89694, Iran}

\begin{abstract}
We characterize phases of the compass ladder model by using degenerate perturbation theory, 
symmetry fractionalization, and numerical techniques. Through degenerate perturbation 
theory we obtain an effective Hamiltonian for each phase of the model, 
and show that a cluster model and the Ising model encapsulate the nature of all phases. 
In particular, the cluster phase has a symmetry-protected topological order, 
protected by a specific $\mathbb{Z}_2\times \mathbb{Z}_2$ symmetry, 
and the Ising phase has a $\mathbb{Z}_2$-symmetry-breaking order characterized 
by a local order parameter
expressed by the magnetization exponent $0.12\pm0.01$.  
The symmetry-protected topological phases inherit all properties of the cluster phases, 
although we show analytically and numerically that they belong to different classes. 
In addition, we study the one-dimensional quantum compass model, 
which naturally emerges from the compass ladder, and show that a partial symmetry 
breaking occurs upon quantum phase transition. 
We numerically demonstrate that a local order parameter accurately determines 
the quantum critical point and its corresponding universality class. 

\end{abstract}
\maketitle

\section{Introduction}
\label{sec:Introduction}

A comprehensive understanding of the phases of matter and also type of transition between them have long been a principal problems in condensed matter physics. It was believed that Landau-Ginzburg theory can help provide such understanding.\cite{book:Goldenfeld} This theory is based on `spontaneous symmetry-breaking phenomenon' associated with a nonzero `local order parameter.' In this theory, all phases of matter are identified by some broken symmetries or, equivalently, by their corresponding local order parameters. However, the emergence of the so-called `topological phases,' which has no evidence of 
symmetry breaking, defies this theory.\cite{Tsui:1982, Kane:2005, Read:1991} Topological phases manifest exotic properties such as robustness against local perturbations \cite{kitaev:2003, Nayak:2008}, nontrivial anyonic statistics\cite{Wen:1990, Wen:2007}, and exhibiting long-range entanglement \cite{Chen:2010}, which make them interesting theoretically and experimentally. 

In the past two decades, vast efforts have been devoted to providing `an alternative framework' 
for characterizing exotic phases of matter. Recently, inspired by ideas from quantum 
information theory (especially distribution of entanglement), ``symmetry fractionalization" 
has been proposed as a technique for full classification of the phases of (quasi) 
one-dimensional ($1$D) gapped quantum systems has been proposed.\cite{Chen:2011:Jan, Chen:2011, Schuch:2011, Pollmann:2010} 
This classification, based on structure of entanglement, 
places the phases into three classes: (i) symmetry-protected topological (SPT) phases, 
which have short-range entanglement, (ii) topologically-trivial phases, 
which can be mapped to fully-product states (with zero entanglement), 
and (iii) symmetry-breaking phases (with degenerate ground states). 
SPT phases, unlike topologically-trivial phases, cannot be mapped to a fully-product state 
as long as 
some specific symmetries are preserved; that is, they are robust against any perturbations which respect these symmetries. 

In symmetry fractionalization, one needs to determine those symmetries which protect a phase, from which a set of unique labels 
are obtained to distinguish the phases that are separated by a quantum phase transition---see Sec.~\ref{sec:symmetryfractionalization}. Obtaining phase labels, however, is a challenging task, which generally requires the prior 
knowledge of symmetries of the model and also an exact infinite matrix product state (iMPS) representation of its ground state. Having determined the symmetries and the iMPS representation of ground state, e.g., by using the infinite time evolving 
block decimation (iTEBD) or infinite-size density matrix renormalization group (iDMRG) methods,\cite{Vidal:2006, Schollwock:2011} one can employ the techniques proposed in Refs.~\onlinecite{Haegeman:2012, Pollmann:2012} to determine phase labels.

There exist numerous 
(exotic) 
models which have been proven to exhibit topological order, but yet a simple and experimentally 
realizable 
model featuring topological phases is of great interest.\cite{Levin:2005,Lin:2014, Fendley:2005} 
In this respect, the Kitaev honeycomb model has been a prominent candidate.\cite{Kitaev:2006:January,Lee:2014,Chaloupka:2013,Reuther:2011,Osorio:2014,Barkeshli:2015} The Hamiltonian of this model contains two-body interactions (hence relatively easier to realize experimentally), and has a rich phase diagram that exhibits different classes of topological phases and non-Abelian anyons. In addition, the Kitaev honeycomb model on an arbitrary-row brick-wall lattice (another representation of the honeycomb lattice) has also been recently studied.\cite{Feng:2007} The associated quantum phase transition between the `exotic phases' of these models are believed to be of topological type, without any (spontaneous) symmetry braking. Nevertheless, the characterization of these phases had remained largely unknown; this is indeed our very goal here to bridge this gap. The model on one- and two-row brick-wall lattices takes a simple form referred to as the ``$1$D compass"\cite{Brzezicki:2007} and the ``compass ladder" models, respectively. Characterization of the corresponding phases is of special importance because these phases (with a proper modification) 
also 
appear in the phase diagram of the Kitaev honeycomb model on arbitrary-row brick-wall lattices. In addition, since ladder systems can be created and manipulated by highly-controlled quantum simulators, they play an important role in experimental realization of `Majorana fermions'---and whence topological quantum computation.\cite{Duan:2003,You:2010, Saket:2010, Tserkovnyak:2011} A promising platform based on the `inhomogeneous Kitaev ladder model' has been recently proposed, which can read out Majorana fermion qubit states and also perform non-Abelian braiding.\cite{He:2013, Pedrocchi:2012, Karimipour:2009, Karimipour:2013, Langari:2015}    

Our main objective in this paper is to identify the type of quantum phase transitions and different topological phases of the compass ladder and $1$D compass models. The compass ladder includes three phases denoted by $\mathfrak{A}$, $\mathfrak{B}$, and $\mathfrak{C}$---see Fig.~\ref{fig:Kitaev phase}. We employ degenerate perturbation theory,\cite{Bergman:2007} to assign an effective Hamiltonian for each phase, which yields: (i) (two different) cluster model(s)\cite{son:2011,Else:2012, Montes:2012}---written in different basis---for the $\mathfrak{A}$ and $\mathfrak{C}$ phases, and (ii) the Ising model for the $\mathfrak{B}$ phase. Based on this analysis, it is shown that the $\mathfrak{A}$ and $\mathfrak{C}$ phases belong to the cluster phase, which is a well-known SPT phase protected by the $\mathbb{Z}_2\times \mathbb{Z}_2$ symmetry. Despite similarity of the $\mathfrak{A}$ and $\mathfrak{C}$ phases, we show that they belong to different classes of SPT phase; the $\mathfrak{A}$ phase is protected by the complex-conjugate symmetry, whereas the $\mathfrak{C}$ phase is not. This observation is also numerically verified by the iTEBD method and the symmetry fractionalization technique. 

The $\mathfrak{B}$ phase appears to be of topologically-trivial $\mathbb{Z}_2$-symmetry-breaking type, characterized by a Landau-type local order parameter. This implies a spontaneous symmetry breaking upon quantum phase transitions, and thus, the phase diagram of the compass ladder can be classified by the associated local order parameter. We demonstrate this result by the iTEBD method after determining the local order parameter and symmetry-breaking group---see Fig.~\ref{fig:Order-Phase-B}. 
In addition, the local order parameter correctly specifies the universality class of the quantum phase transitions as of the Ising class (with the magnetization exponent $\beta=1/8$). We remark that our conclusion differs with Ref.~\onlinecite{Feng:2007}, where the classification of the phase diagram is based on nonlocal string order parameters (whereby believed that there were no explicit change of symmetry upon quantum phase transitions). 

Additionally, we study the $1$D compass model, which naturally appears by turning off one of the 
coupling parameters of the compass ladder. Upon quantum phase transition 
a specific $\mathbb{Z}_{2}$ symmetry is broken and another one is preserved, 
thus a partial spontaneous symmetry breaking occurs
---i.e. a quantum phase transition between two phases, 
where in each phase, part of symmetry group has been broken. 
Based on this fact, one can construct a local order parameter to capture quantum 
the phase transitions and relevant physics of the model. 
Interestingly, as examined by the iTEBD method, this local order parameter is shown to give the accurate values of both critical point and magnetization exponent ($\beta=1/8$).       

This paper is organized as follows. In Sec.~\ref{sec:modle} the models and their phase diagram are reviewed. In Sec.~\ref{sec:perturbationtheory} the effective Hamiltonian of the compass ladder is obtained. Broken symmetry of the $\mathfrak{B}$ phase and its corresponding local order parameter are derived in Sec.~\ref{sec:characterization}, and numerically examined. The implementation of the symmetry fractionalization technique to obtain the labels of the SPT phases are presented in Sec.~\ref{sec:symmetryfractionalization}, and the topological properties of the SPT phases are discussed next in Sec.~\ref{sec:topologicalorder}. We discuss the phase characterization of the compass model in Sec.~\ref{sec:compassmodel}.
The paper is concluded in Sec.~\ref{sec:conclusion} with a summary of our results.   

\section{Compass Ladder model}
\label{sec:modle}

\begin{figure}
\includegraphics[width=1.0 \linewidth]{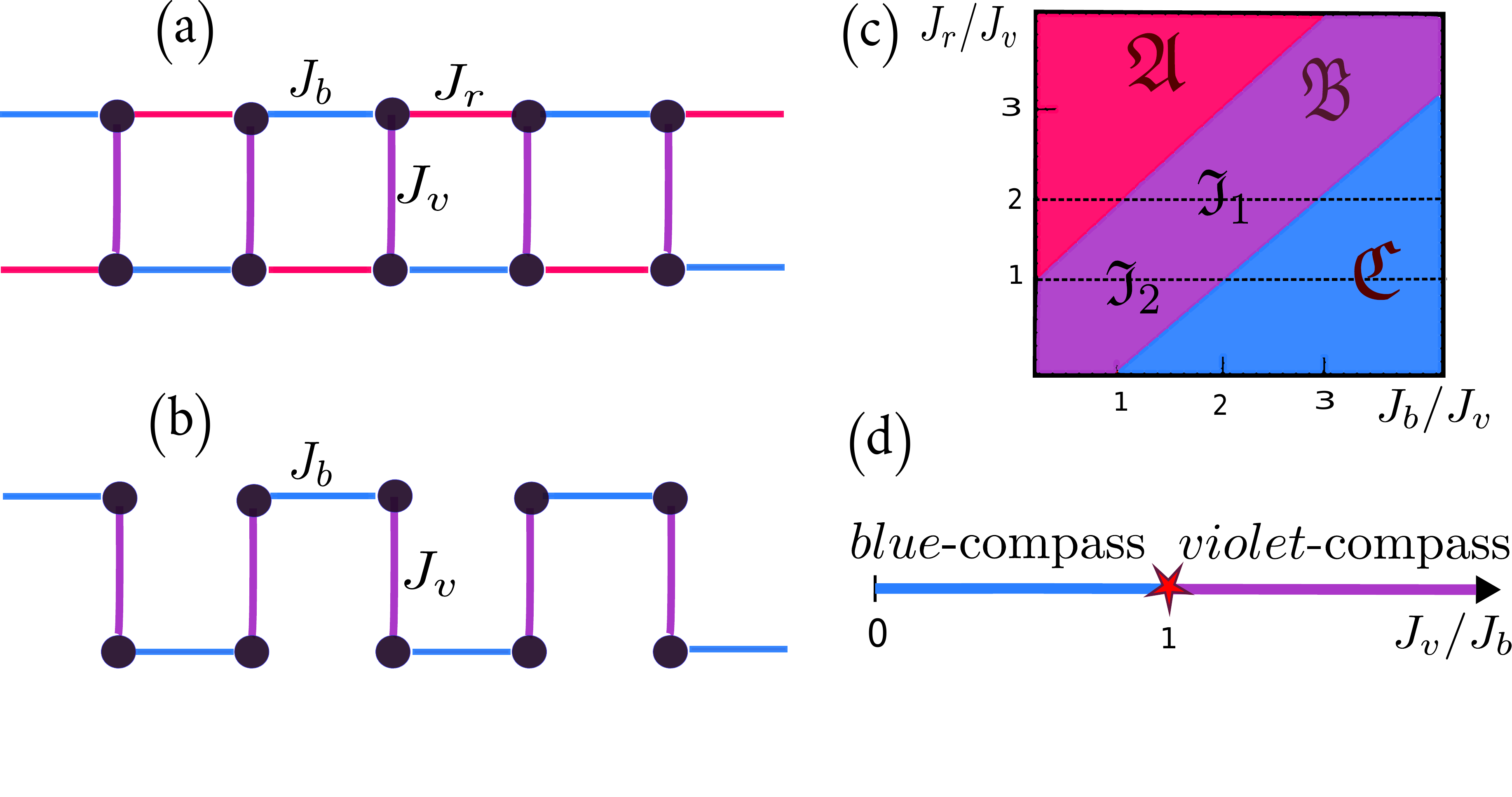}
\caption{(Color online). Graphical representations and phase diagrams of the compass ladder and $1$D compass models. Black circles and colored links represent spin-$1/2$ particles and different types of interactions, respectively. (a) The compass ladder model. (b) The $1$D compass model, obtained by switching off the red-link interactions of the compass ladder in (a). (c) The phase diagram of the compass ladder model. Here the two paths $\mathfrak{I}_{1}$ and $\mathfrak{I}_{2}$ are introduced for our numerical analysis. (d) The phase diagram of $1$D compass model.}
\label{fig:Kitaev phase}
\end{figure}

The compass ladder model (also referred to as the XYZ compass model \cite{Nussinov:2015}) is defined on a ladder geometry as in Fig.~\ref{fig:Kitaev phase}-(a), where the black circles denote spin-$1/2$ particles, and the colored links (blue, red, and violet) 
represent different types of interaction denoted, respectively, by `$\mathrm{b}$,' `$\mathrm{v}$,' and `$\mathrm{r}$.' The Hamiltonian is given by
\begin{equation}
H_{\mathrm{KL}}=-J_{\mathrm{b}} \sum_{\mathrm{b~links}}\sigma_{i}^{x}\sigma_{j}^{x}-J_{\mathrm{r}}\sum_{\mathrm{r~links}} \sigma_{i}^{y}\sigma_{j}^{y}-J_{\mathrm{v}}\sum_{\mathrm{v~links}}\sigma_{i}^{z}\sigma_{j}^{z},
\end{equation}
where $\sigma^{\alpha}$ (for $\alpha\in\{x,y,z\}$) represents the $\alpha$ Pauli matrix, and $J_{\mathrm{a}}$ (for  $\mathrm{a}\in\{\mathrm{r},\mathrm{v},\mathrm{b}\}$) is the coupling constant. Without loss of generality, the coupling constants are assumed to be positive; $J_{\mathrm{a}}\geqslant0$. In Ref.~\onlinecite{Feng:2007} the phase diagram of the model has been obtained as in Fig.~\ref{fig:Kitaev phase}-(c) through the Jordan-Wigner transformation technique. This diagram contains three gapped phases labelled by $\mathfrak{A}$, $\mathfrak{B}$, and $\mathfrak{C}$. The $\mathfrak{A}$ ($\mathfrak{B}$) phase is separated from the $\mathfrak{B}$ ($\mathfrak{C}$) phase by the gapless line $J_{\mathrm{r}}/J_{\mathrm{v}}=J_{\mathrm{b}}/J_{\mathrm{v}}+1$ ($J_{\mathrm{r}}/J_{\mathrm{v}}=J_{\mathrm{b}}/J_{\mathrm{v}}-1$). The quantum phase transition between these phases is of the second-order type (because of the divergence in the second derivative of the ground-state energy), and was believed to be topological (characterized by string order parameters).

The compass ladder model reduces to the $1$D compass model when one of the coupling constants vanishes. For the case of $J_{\mathrm{r}}=0$, as shown in Fig.~\ref{fig:Kitaev phase}-(b), the Hamiltonian reduces to
\begin{equation}
H_{\mathrm{compass}}=-J_{\mathrm{b}} \sum_{\mathrm{b~links}}\sigma_{i}^{x}\sigma_{j}^{x}-J_{\mathrm{v}}\sum_{\mathrm{v~links}}\sigma_{i}^{z}\sigma_{j}^{z}.
\label{Eq:CompassModel}
\end{equation}
The phase diagram of the $1$D compass model is already known as in Fig.~\ref{fig:Kitaev phase}-(d), and contains two gapped phases with extensive degeneracy separated at the critical point $J_{\mathrm{v}}/J_{\mathrm{b}}=1$. The gapped phases are called the blue- and violet-compass phases for $J_{\mathrm{v}}/J_{\mathrm{b}}<1$ and $J_{\mathrm{v}}/J_{\mathrm{b}}>1$, respectively. The quantum phase transition is topological and of the second-order type. Similar to the compass ladder, the nature of the topological quantum phase transition in the $1$D compass model has been shown through nonlocal string order parameters.\cite{Wang:2015} 

To identify the nature of the $\mathfrak{A}$, $\mathfrak{B}$, and $\mathfrak{C}$ phases and their corresponding quantum phase transitions, we derive the effective Hamiltonian\cite{kargarian:2010} of the compass ladder associated with each phase, which capture main physical properties of each phase. 

\section{Degenerate perturbation theory and effective Hamiltonians}
\label{sec:perturbationtheory}

Degenerate perturbation theory is based on splitting the original Hamiltonian $H$ into two parts: $H_{0}$ and $V$. The $H_{0}$ term represents the unperturbed Hamiltonian, whose energy spectrum is fully known, and in general could be degenerate. The $V$ term plays the role of perturbation, whose operator norm is relatively smaller than the spectral gap $\Delta_0$ of $H_0$, i.e., $\Vert V\Vert\ll\Delta_0$. In the case of the degenerate perturbation formalism, for a specific energy level, the correction at the $m$th order of perturbation is given by an `effective Hamiltonian.' For a quantum phase transition, the effective Hamiltonian for ground-state energy is required, which is denoted by $H^{(m)}_{\mathrm{eff}}$.\cite{Bergman:2007}
  
The starting point to obtain $H^{(m)}_{\mathrm{eff}}$ is to define the projection operator into the `unperturbed degenerate ground space' (set of all ground states of $H_{0}$),
\begin{equation}
\mathpzc{P}=\sum_{i:~H_{0}|\Psi^{i}_{0}\rangle=E_{0}|\Psi^{i}_{0}\rangle}|\Psi^{i}_{0}\rangle \langle\Psi^{i}_{0}|, 
\end{equation}  
where $E_{0}$ is the ground-state energy of $H_{0}$. Having determined $\mathpzc{P}$, the effective Hamiltonian $H^{(m)}_{\mathrm{eff}}$ can be determined. The first-order effective Hamiltonian $H^{(1)}_{\mathrm{eff}}$ has the following form:
\begin{equation}
H^{(1)}_{\mathrm{eff}}=\mathpzc{P}V\mathpzc{P}. 
\end{equation}  
The form of higher orders of the effective Hamiltonian becomes gradually more complex,; e.g., the second- and third-order effective Hamiltonians are given by 
\begin{align}
H^{(2)}_{\mathrm{eff}} &=\mathpzc{P}VGV\mathpzc{P}, \\
H^{(3)}_{\mathrm{eff}} &=\mathpzc{P}VGVGV\mathpzc{P}-E^{(1)}_{0}\mathpzc{P}VGGV\mathpzc{P}, 
\end{align}  
where
\begin{equation}
G=\frac{1}{E_{0}-H_{0}}(\openone-\mathpzc{P})
\end{equation}
is the Green's function, and $E^{(1)}_{0}$ denotes the ground-state energy of $H^{(1)}_{\mathrm{eff}}$. 

\subsection{Effective Hamiltonian associated with the $\mathfrak{B}$ phase}
\label{Sec:PhaseB}

To obtain the effective Hamiltonian for the $\mathfrak{B}$ phase, $H_{0}$ and $V$ are set as follows:
\begin{align*}
H_{0}=&J_{\mathrm{v}}\sum_{\mathrm{v~links}}\sigma_{i}^{z}\sigma_{j}^{z}, \\
 V=&J_{\mathrm{b}} \sum_{\mathrm{b~links}}\sigma_{i}^{x}\sigma_{j}^{x}+J_{\mathrm{r}}\sum_{\mathrm{r~links}} \sigma_{i}^{y}\sigma_{j}^{y},
\end{align*}
where $J_{\mathrm{r}}, J_{\mathrm{b}}\ll J_{\mathrm{v}}$ (note that positivity of $J_{\mathrm{v}}$ and nonzero values of $J_{\mathrm{r}}$ and $ J_{\mathrm{b}}$ guarantee that the ground state of $H=H_{0}+V$ is within the $\mathfrak{B}$ phase (see Fig.~\ref{fig:Kitaev phase}-(c)). 

The projection operator $\mathpzc{P}_{\mathrm{v}}$, which comes from the unperturbed degenerate ground space (set of all highly-degenerate ground states of $H_{0}$), is defined as follows:
\begin{equation}
\mathpzc{P}_{\mathrm{v}}=\prod_{\mathrm{v~links}} \mathpzc{P}_{\mathrm{v}}^{0},\quad \mathpzc{P}_{\mathrm{v}}^{0}=|\uparrow\uparrow\rangle\langle\uparrow\uparrow|+|\downarrow\downarrow\rangle\langle\downarrow\downarrow|,
\end{equation}  
where $|\uparrow\rangle$ and $|\downarrow\rangle$ are the eigenstates of $\sigma^{z}$---index $\mathrm{v}$ denotes violet. 
However, for simplicity it is more convenient to write $H_{\mathrm{eff}}$ in a new basis by rewriting  $\mathpzc{P}_{0}$ as
\begin{equation}
\mathpzc{P}_{\mathrm{v}}^{0}=|\overline{\uparrow}\rangle\langle\uparrow\uparrow|+|\overline{\downarrow}\rangle\langle\downarrow\downarrow|,
\end{equation}
where $|\overline{\uparrow}\rangle\equiv |\uparrow\uparrow\rangle$ and $|\overline{\downarrow}\rangle \equiv |\downarrow\downarrow\rangle$ 
are the `logical qubits' in the $\sigma^{z}$-basis. The energy and degeneracy of an unperturbed ground state are, respectively, equal to $E_{0}=-N_{\mathrm{v}}J_{\mathrm{v}}$ and $2^{N_{\mathrm{v}}}$, where $N_{\mathrm{v}}$ is number of the violet links. 
The first excitation of $H_{0}$ has energy $E_{1}=-(N_{\mathrm{v}}-2)J_{\mathrm{v}}$ with degeneracy $2N_{\mathrm{v}}2^{N_{\mathrm{v}}-1}$, which is  obtained by flipping one of the spins. Flipping two spins on different violet links gives rise to higher 
exited states that has the energy $E_{2}=-(N_{\mathrm{v}}-4)J_{\mathrm{v}}$ with degeneracy $2(N_{\mathrm{v}}-1)N_{\mathrm{v}}2^{N_{\mathrm{v}}-2}$.

\begin{figure}
\includegraphics[width=1.0 \linewidth]{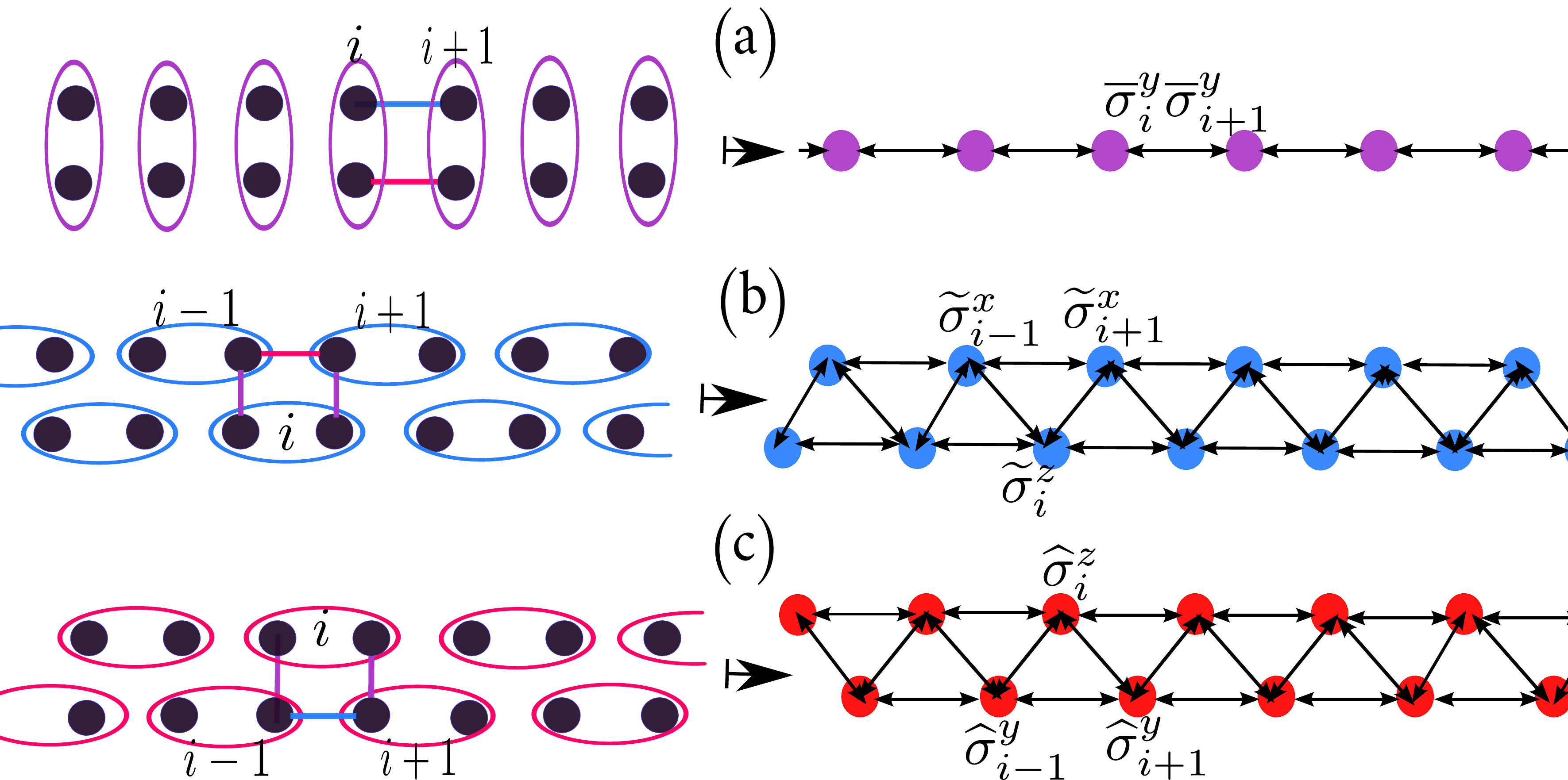}
\caption{(Color online). Schematic representation of the effective Hamiltonians of the $\mathfrak{A}$, $\mathfrak{B}$, and $\mathfrak{C}$ phases. The vertical violet ellipses show that the state of the two spins (denoted by the black circles placed within the ellipses) is either $|\uparrow\uparrow\rangle$ or $|\downarrow\downarrow\rangle$. The horizontal blue and red ellipses also represent the states $\{ |\nearrow\nearrow\rangle, |\swarrow\swarrow\rangle\}$ and $\{ |\nwarrow\nwarrow\rangle, |\searrow\searrow\rangle\}$, respectively (see the main text). Colored links and circles denote different types of intersections and logical qubits. Index $i$ labels the ellipses [left], or equivalently logical qubits [right]. (a), (b), and (c) [left]: The main contributions to the effective Hamiltonians of the $\mathfrak{B}$, $\mathfrak{C}$, and $\mathfrak{A}$ phases, respectively, and [right] 
their corresponding effective Hamiltonians. }
\label{fig:Perturbation}
\end{figure}

The first-order effective Hamiltonian ($\mathpzc{P}V\mathpzc{P}$) is zero because $V$ excites the unperturbed ground space into the second-excited subspace, which obviously has no overlap with the unperturbed ground-state subspace; whence $H^{(1)}_{\mathrm{eff}}=0$. However, the second-order effective Hamiltonian is nonzero, resulting in both nontrivial and trivial terms, (nontrivial terms break the highly-degenerate ground-state subspace, while the trivial terms do not). In the expression $\mathpzc{P}VGV\mathpzc{P}$, one of the possibilities (among many) is to choose the first and second $V$ on a specific link. The first $V$ excites unperturbed ground space, bringing it to the second excited space. The effect of the Green's function on the second excited space is $G=-1/(4J_{\mathrm{v}})$, and the second $V$ takes the second excited space back to the unperturbed ground space. Thus, one can show that such interactions yield trivial contributions to the second-order effective Hamiltonian
\begin{equation}
H^{(2)}_{\mathrm{eff}}= -N_{\mathrm{v}}\frac{J^{2}_{\mathrm{r}}}{4J_{\mathrm{v}}} \prod_{i} \overline{I}_{i}-N_{\mathrm{v}}\frac{J^{2}_{\mathrm{b}}}{4J_{\mathrm{v}}} \prod_{i} \overline{I}_{i},
\end{equation} 
where $\overline{I}=|\overline{\uparrow}\rangle \langle \overline{\uparrow}|+|\overline{\downarrow}\rangle \langle \overline{\downarrow}|$ and $i$ runs over the violet links, see Fig.~\ref{fig:Perturbation}-(a) [left].

The other nonzero contributions to the second-order effective Hamiltonian $H^{(2)}_{\mathrm{eff}}$ are from those interactions that act on nearest neighbor violet links, as sketched in Fig.~\ref{fig:Perturbation}-(a) [left]. In this case, the first $V$---the blue link in Fig.~\ref{fig:Perturbation}-(a) [left]---excites the unperturbed ground space, resulting in the second excited space. The action of the Green's function on this second excited space is given by $G=-1/(4J_{\mathrm{v}})$. The second $V$---the red link in Fig.~\ref{fig:Perturbation}-(a) [left]---takes the second excited space back into the unperturbed ground space. It is straightforward to show that $\mathpzc{P}VGV\mathpzc{P}$ for Fig.~\ref{fig:Perturbation}-(a) [left] is proportional to $\cdots\otimes\overline{I}\otimes\overline{\sigma}^{y}\otimes \overline{\sigma}^{y}\otimes\overline{I}\otimes\cdots$, where $\overline{\sigma}^{y}$ is the $y$ Pauli matrix in the logical basis $\{|\overline{\uparrow}\rangle,|\overline{\downarrow}\rangle\}$, i.e.,
\begin{equation}
\overline{\sigma}^{y}=-i|\overline{\uparrow}\rangle\langle\overline{\downarrow}|+i|\overline{\downarrow}\rangle\langle\overline{\uparrow}|.
\end{equation}
We note that the terms acting on the next-nearest-neighbor violet links (or farther neighbors) play no role in the second-order effective Hamiltonian. 

In summary, $H^{(2)}_{\mathrm{eff}}$ is given by, as shown in Fig.~\ref{fig:Perturbation}-(a) [right], 
\begin{align}
H^{(2)}_{\mathrm{eff}}&=-N\frac{J^{2}_{\mathrm{r}}}{4J_{\mathrm{v}}} \prod_{i} \overline{I}_{i}-N\frac{J^{2}_{\mathrm{b}}}{4J_{\mathrm{v}}} \prod_{i} \overline{I}_{i}-2\frac{J_{\mathrm{b}}J_{\mathrm{r}}}{4J_{\mathrm{v}}}\sum_{i}\overline{\sigma}_{i}^{y} \, \overline{\sigma}_{i+1}^{y}\nonumber \\
 &=-2\frac{J_{\mathrm{b}}J_{\mathrm{r}}}{4J_{\mathrm{v}}}\sum_{i}\overline{\sigma}_{i}^{y} \; \overline{\sigma}_{i+1}^{y}+\mathrm{const.}
\label{Ham:Ising}
\end{align}
The factor $2$ indicates that there are two possibilities for choosing the first and second $V$ in 
the expression $\mathpzc{P}VGV\mathpzc{P}$. 

\subsection{Effective Hamiltonian associated with the $\mathfrak{C}$ phase}
\label{Sec:PhaseC}

Here $H_{0}$ and $V$ are defined as follows:
\begin{align}
H_{0}&=J_{\mathrm{b}}\sum_{\mathrm{b~links}}\sigma_{i}^{x}\sigma_{j}^{x}, \\
 V&=J_{\mathrm{r}} \sum_{\mathrm{r~links}}\sigma_{i}^{y}\sigma_{j}^{y}+J_{\mathrm{v}}\sum_{\mathrm{v~links}} \sigma_{i}^{z}\sigma_{j}^{z}.
\end{align}
where $J_{\mathrm{v}}, J_{\mathrm{b}}\ll J_{\mathrm{r}}$. This sort of definition of $H_{0}$, $V$, and the coupling constants is to guarantee that the ground state (of $H$) is placed within the $\mathfrak{C}$ phase. Similar to Sec.~\ref{Sec:PhaseB}, the goal is to obtain the leading-order nontrivial effective Hamiltonian.     

The projection operator into the highly-degenerate ground state of $H_{0}$ is given as follows:
\begin{equation*}
\mathpzc{P}_{\mathrm{b}}=\prod_{\mathrm{b~links}} \mathpzc{P}_{\mathrm{b}}^{0},\quad \mathpzc{P}_{\mathrm{b}}^{0}\equiv|\nearrow\nearrow\rangle\langle\nearrow\nearrow|+|\swarrow\swarrow\rangle\langle\swarrow\swarrow|,
\end{equation*}  
where $|\nearrow\rangle$ and $|\searrow\rangle$ are the eigenstates of $\sigma^{x}$. Rewriting $\mathpzc{P}_{\mathrm{b}}^{0}$ in a new basis makes the form of the effective Hamiltonian simpler as
\begin{equation}
\mathpzc{P}_{\mathrm{b}}^{0}=|\widetilde{\nearrow}\rangle\langle\nearrow\nearrow|+|\widetilde{\swarrow}\rangle\langle\swarrow\swarrow|,
\end{equation}
where $|\widetilde{\nearrow}\rangle=|\nearrow\nearrow\rangle$ and $|\widetilde{\swarrow}\rangle=|\swarrow\swarrow\rangle$ are the logical qubits in the $\sigma^{x}$-basis. The energy and number of degeneracy of the unperturbed ground space are the same as Sec.~\ref{Sec:PhaseB}, i.e., $E_{0}=-N_{\mathrm{b}}J_{\mathrm{b}}$ and $2^{N_{\mathrm{b}}}$, where $N_{\mathrm{b}}$ is number of the blue links. The first (second) unperturbed excited space is obtained by flipping one (two) spin(s) on a specific (two different) blue link(s), which give $E_{1}=-(N_{\mathrm{b}}-2)J_{\mathrm{b}}$ ( $E_{2}=-(N_{\mathrm{b}}-4)J_{\mathrm{b}}$) with degeneracy $2N_{\mathrm{b}}2^{N_{\mathrm{b}}-1}$ ($2(N_{\mathrm{b}}-1)N_{\mathrm{b}}2^{N_{\mathrm{b}}-2}$).
  
Similar to Sec.~\ref{Sec:PhaseB}, the first-order effective Hamiltonian is zero: $H^{(1)}_{\mathrm{eff}}=\mathpzc{P}_{\mathrm{b}}V\mathpzc{P}_{\mathrm{b}}=0$. The second-order effective Hamiltonian results in trivial terms: the only possibility to have nonzero terms for $\mathpzc{P}VGV\mathpzc{P}$ is to choose the first and the second $V$ on a specific link. It yields  
\begin{equation}
H^{(2)}_{\mathrm{eff}}=-N_{\mathrm{b}}\frac{J^{2}_{\mathrm{v}}}{4J_{\mathrm{b}}} \prod_{i} \widetilde{I_{i}}-N_{\mathrm{b}}\frac{J^{2}_{\mathrm{r}}}{4J_{\mathrm{b}}} \prod_{i} \widetilde{I_{i}}, 
\end{equation}       
where $\widetilde{I}=|\widetilde{\nearrow}\rangle\langle\widetilde{\nearrow}|+|\widetilde{\searrow}\rangle\langle\widetilde{\searrow}|$, 
and $i$ runs over logical qubits, as shown in Fig.~\ref{fig:Perturbation}-(b). 
The third-order effective Hamiltonian, 
$H^{(3)}_{\mathrm{eff}}=\mathpzc{P}_{\mathrm{b}}VGVGV\mathpzc{P}_{\mathrm{b}}-E^{(1)}_{0}\mathpzc{P}_{\mathrm{b}}VGG\mathpzc{P}_{\mathrm{b}}$, 
leads to a nontrivial term. The second term of $H^{(3)}_{\mathrm{eff}}$ vanishes because $E_{0}^{(1)}=0$. 
The closed form of the first term ($\mathpzc{P}_{\mathrm{b}}VGVGV\mathpzc{P}_{\mathrm{b}}$) is obtained 
by such choices as depicted in Fig.~\ref{fig:Perturbation}-(b) [left]. Suppose the first and 
the second $V$ are the violet-link interactions, and the third $V$ is the red-link one. 
The first $V$ excites the unperturbed ground space to the second excited space. 
The effect of the Green's function $G$ on the the second excited space is $G=-1/(4J_{\mathrm{b}})$. 
The second $V$ just transforms the second excited state to itself; that is, 
the second $V$ only rotates the states within the second excited space. 
Thus, when the next $G$ is applied, $G=-1/(4J_{\mathrm{b}})$. 
The third $V$ takes the second excited state back into the unperturbed ground space. 
It can be shown that the expression $\mathpzc{P}_{\mathrm{b}}VGVGV\mathpzc{P}_{\mathrm{b}}$, in Fig.~\ref{fig:Perturbation}-(b) [left], is proportional to $\cdots\otimes\widetilde{I}\otimes\widetilde{\sigma}^{x}\otimes\widetilde{\sigma}^{z}\otimes\widetilde{\sigma}^{x}\otimes\widetilde{I}\otimes\cdots$, where $\widetilde{\sigma}^{x}$ and $\widetilde{\sigma}^{z}$ are the $x$ and $z$ Pauli matrices in the logical basis $\{|\widetilde{\nearrow} \rangle ,|\widetilde{\swarrow} \rangle \}$. Here $\widetilde{\sigma}^{x}$ and $\widetilde{\sigma}^{z}$ are given by
\begin{equation}
\widetilde{\sigma}^{x}=|\widetilde{\nearrow} \rangle \langle\widetilde{\nearrow}|-|\widetilde{\swarrow} \rangle \langle\widetilde{\swarrow}|,
\: \widetilde{\sigma}^{z}=|\widetilde{\nearrow} \rangle \langle\widetilde{\swarrow}|+|\widetilde{\swarrow} \rangle \langle\widetilde{\nearrow}|.
\end{equation}
Other selections
of $\mathpzc{P}_{\mathrm{b}}VGVGV\mathpzc{P}_{\mathrm{b}}$---except those in Fig.~\ref{fig:Perturbation}-(b) [left]---make no contribution to $H^{(3)}_{\mathrm{eff}}$, whence
\begin{equation}
H^{(3)}_{\mathrm{eff}}=-2\frac{J_{\mathrm{r}}J_{\mathrm{v}}^{2}}{{(4J_{\mathrm{b}})}^{2}}\sum_{i}\widetilde{\sigma}_{i-1}^{x}
\widetilde{\sigma}_{i}^{z}\widetilde{\sigma}_{i+1}^{x}. 
\label{Ham:ClusterX}
\end{equation}
The factor $2$ is again due to different choices of $V$---there are $6$ different configurations, similar to that of Fig.~\ref{fig:Perturbation}-(b) [left], whose factors cancel out each other as $-2+2-2=-2$. Equation~(\ref{Ham:ClusterX}) is the cluster Hamiltonian, which belongs to the class of stabilizer Hamiltonians. 
The ground state of the cluster Hamiltonian has a unique (for periodic boundary condition) and exact MPS form, and is of the $\mathbb{Z}_2\times\mathbb{Z}_2$ SPT type.\cite{Else:2012} 
 
\subsection{Effective Hamiltonian associated with the $\mathfrak{A}$ phase}
\label{Sec:PhaseA}

The effective Hamiltonian of the $\mathfrak{A}$ phase can be obtained by replacing $J_{\mathrm{r}} \rightarrow J_{\mathrm{b}}$ and $\sigma^x (\widetilde{\sigma}^{x}) \rightarrow \sigma^y (\widetilde{\sigma}^{y})$ in the results of Sec.~\ref{Sec:PhaseC}, which yields
\begin{equation}
H^{(3)}_{\mathrm{eff}}=-2\frac{J_{\mathrm{b}}J_{\mathrm{v}}^{2}}{{(4J_{\mathrm{r}})}^{2}}\sum_{i}\widehat{\sigma}_{i-1}^{y}
\widehat{\sigma}_{i}^{z}\widehat{\sigma}_{i+1}^{y}, 
\label{Ham:ClusterY}
\end{equation}
where $\widehat{\sigma}^{y}$ and $\widehat{\sigma}^{z}$ are the $y$ and $z$ Pauli matrices in the logical basis $\{|\widehat{\nwarrow} \rangle ,|\widehat{\searrow} \rangle \}$. In this basis, 
\begin{equation}
|\widehat{\nwarrow} \rangle=|\nwarrow\nwarrow\rangle,\quad |\widehat{\searrow} \rangle=|\searrow\searrow\rangle,
\end{equation}
where $|\nwarrow\rangle$ and $|\searrow\rangle$ are the eigenvectors of $\sigma^{y}$. Equation~(\ref{Ham:ClusterY}) is the cluster Hamiltonian written in a different basis; it can be obtained from Eq.~(\ref{Ham:ClusterX}) by $\pi/2$-rotation about the $z$-axis. Since this operation is unitary, the ground state of the Hamiltonian (\ref{Ham:ClusterY}) inherits the properties of the cluster phase such as having unique exact MPS form and being of the $\mathbb{Z}_2\times\mathbb{Z}_2$ SPT type.    

\section{Characterization of different phases}
\label{sec:characterization}

\subsection{Infinite matrix product state (iMPS) method}
\label{sec:iMPS}

\begin{figure}
\includegraphics[width=1.0 \linewidth]{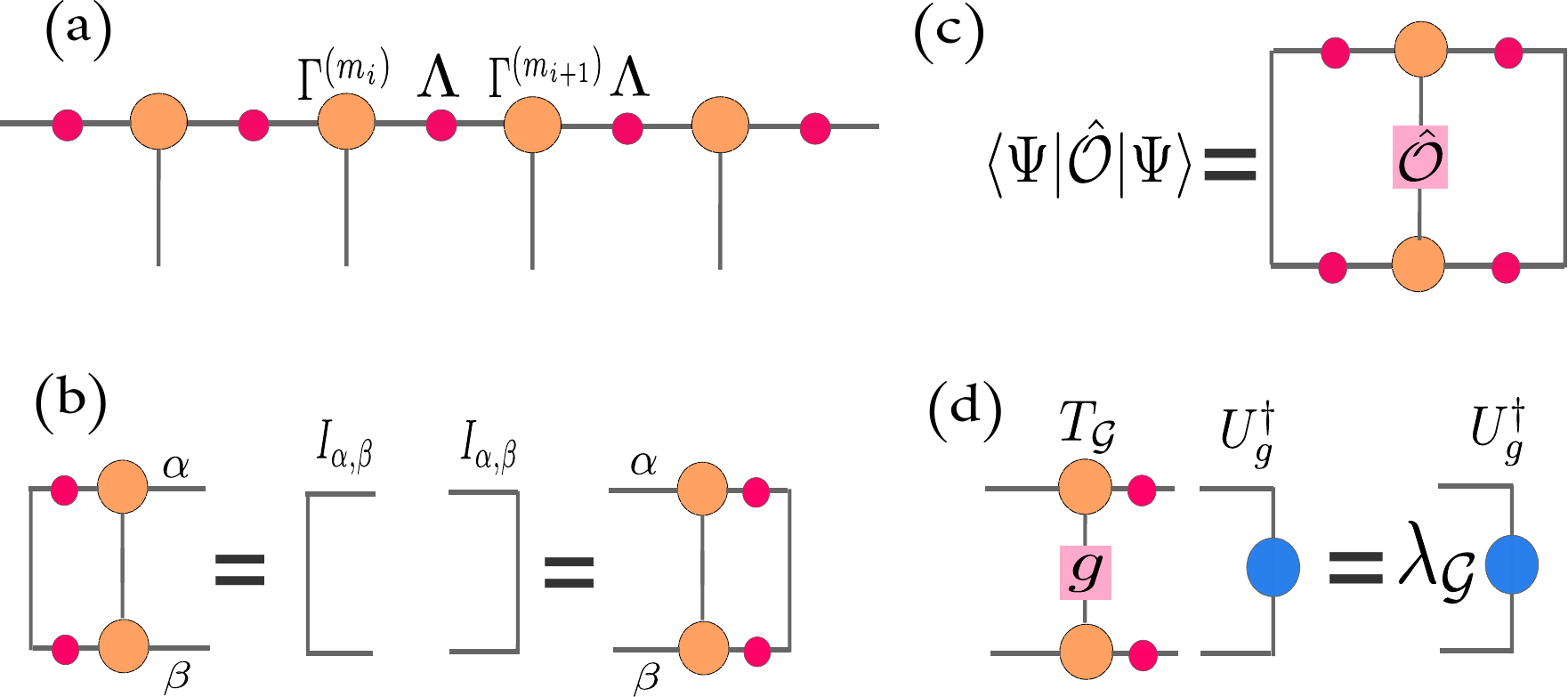}
\caption{(Color online). Diagrammatic representation of iMPS and some related quantities. 
(a) Graphical representation of Eq.~(\ref{Eq:iMPS}) with the matrices $(\Gamma^{(m_{i})}, \Lambda)$. 
(b) Conditions for canonical iMPS. (c) Expectation value of an on-site operator $\hat{{\mathpzc{O}}}$ 
in the canonical iMPS form. (d) Eigenvalue equation of the $\mathpzc{G}$-transfer matrix 
with its maximum eigenvalue ($\lambda_{\mathpzc{G}}$) and the corresponding right 
eigenstate ($U^{\dagger}_{g}$). If $\lambda_{\mathpzc{G}}=1$, the iMPS is symmetric under $\mathpzc{G}$.}
\label{fig:iMPS}
\end{figure}

Ground state of (quasi) $1$D gapped quantum systems respects `area law,' in the sense that bipartite entanglement of an arbitrary subsystem depends on its boundary rather than bulk. Based on this fact, it has been proven that (quasi) $1$D gapped 
quantum phases can be faithfully represented by iMPSs. \cite{Hastings:2007} The iMPS representation of a state $|\Psi\rangle$ (ground state of a $1$D gapped system) is based on assigning to each site a set of matrices as
\begin{equation}
|\Psi\rangle=\sum_{\cdots \scalebox{0.64}{$m_{i},m_{i+1}$} \cdots}  \cdots \Gamma^{(m_{i})} \Lambda \Gamma^{(m_{i+1})} \Lambda\cdots |\cdots \scalebox{0.84}{$(m_{i})(m_{i+1})$} \cdots\rangle,
\label{Eq:iMPS}
\end{equation}  
where $\Lambda$ is a $D\times D$ diagonal matrix, and $\Gamma^{(m_{i})}$s are some $D\times D$ matrices assigned to site $i$ [Fig.~\ref{fig:iMPS}-(a)]. The matrices $(\Gamma^{(m_{i})}, \Lambda)$ are usually determined by the iTEBD or iDMRG methods, where the accuracy of the scheme is controlled by the parameter $D$. Having determined the matrices $(\Gamma^{(m_{i})}, \Lambda)$, one can always use a `canonical transformation' and rewrite the iMPS representation in a more suitable \textit{canonical} form: $(\Gamma^{(m_{i})}, \Lambda)\rightsquigarrow (\widetilde{\Gamma}^{(m_{i})}, \widetilde{\Lambda})$ \cite{Orus:2008}. In the canonical iMPS form, as shown in Fig.~\ref{fig:iMPS}-(b), new matrices $(\widetilde{\Gamma}^{(m_{i})}, \widetilde{\Lambda})$ satisfy the following conditions:
\begin{align}
&\sum_{m_{i}} (\widetilde{\Gamma}^{(m_{i})} \widetilde{\Lambda})  (\widetilde{\Gamma}^{(m_{i})} \widetilde{\Lambda})^{\dagger} =\openone,\\
&\sum_{m_{i}} ({\widetilde{\Lambda}\widetilde{\Gamma}^{(m_{i})}})^{\dagger}  (\widetilde{\Lambda}\widetilde{\Gamma}^{(m_{i})})   =\openone.
\end{align}
where $\widetilde{\Lambda}$ is a positive diagonal matrix related to the density matrix of a half of the system through $\varrho=\widetilde{\Lambda}^{2}$. In this form, the expectation value of a local order parameter (defined on a given site) is given by
\begin{align}
\langle\Psi| \hat{{\mathpzc{O}}} |\Psi\rangle =& \sum_{m_{i} m'_{i} \alpha\beta} (\widetilde{\Lambda}_{(\alpha)}\widetilde{\Gamma}_{(\alpha)(\beta)}^{(m_{i})}
\widetilde{\Lambda}_{(\beta)}) \hat{{\mathpzc{O}}}_{(m_{i}), (m'_{i})}\nonumber\\
&~\times (\widetilde{\Lambda}_{(\alpha)}\widetilde{\Gamma}_{(\alpha)(\beta)}^{(m'_{i})}  \widetilde{\Lambda}_{(\beta)})^{\ast},
\end{align}
as depicted in Fig.~\ref{fig:iMPS}-(c).
In addition, the on-site symmetry groups can be evaluated in a straightforward manner in the canonical iMPS representation of the ground state $|\Psi\rangle$. The on-site symmetry $\mathpzc{G}=\prod_{i=1}^{N}g_{i}$ is respected by $|\Psi\rangle$ if in the following relation $\lambda_\mathpzc{G}$ becomes $1$:
\begin{equation}
\lim_{N \rightarrow +\infty}\langle\Psi| \prod_{i=1}^{N}g_{i} |\Psi\rangle=\lim_{N \rightarrow +\infty}\mathrm{Tr}[T^{N}_{\mathpzc{G}}]=\lambda_{\mathpzc{G}}^{N}, 
\end{equation} 
where $T_{\mathpzc{G}}$ is the $\mathpzc{G}$-transfer matrix 
\begin{equation}
{T_{\mathpzc{G}}}_{(\alpha\alpha'),(\beta\beta')} = \sum_{m_{i}m'_{i}} (\widetilde{\Gamma}_{\alpha\beta}^{(m_{i})}
\widetilde{\Lambda}_{\beta}) (g_{(m_{i}),(m'_{i})}) (\widetilde{\Gamma}_{\alpha'\beta'}^{m'_{i}}  \widetilde{\Lambda}_{\beta'})^{\ast},
\end{equation} 
shown in Fig.~\ref{fig:iMPS}-(d), and $\lambda_\mathpzc{G}$ is its maximum eigenvalue. Furthermore, if the symmetry $\mathpzc{G}$ is respected (that is, $\lambda_\mathpzc{G}=1$), the following relation should be satisfied:
\begin{equation}
 \sum_{m'_{i}} g_{(m_{i}),(m'_{i})} \widetilde{\Gamma}^{(m'_{i})} = e^{i\theta_{g}}
U^{\dagger}_{g} \widetilde{\Gamma}^{(m_{i})} U_{g},
\label{Eq:SymmetryPreservation}
\end{equation}
where $e^{i\theta_{g}}$ is a phase, and $U_{g}$ is a unitary matrix (which plays an important 
role in the classification of SPT phases)---see Sec.~\ref{sec:symmetryfractionalization}. 
It is straightforward to show that the right eigenstate of 
the $\mathpzc{G}$-transfer matrix $T_{\mathpzc{G}}$ (corresponding to the eigenvalue $\lambda_\mathpzc{G}$) 
is $U^{\dagger}_{g}$ (see Fig.~\ref{fig:iMPS}-(d)).

\subsection{Local order parameter}
\label{sec:loacalorder}

The nature of the $\mathfrak{B}$ phase is revealed by the Ising Hamiltonian (\ref{Ham:Ising}). This Hamiltonian has two fully-product degenerate ground states, implying that the $\mathfrak{B}$ phase is of the topologically-trivial $\mathbb{Z}_2$-symmetry-breaking type. The $\mathbb{Z}_2$-symmetry-broken group and the corresponding local order parameter ($\overline{\mathpzc{O}}$), in the logical basis, are given by
\begin{equation}
\mathbb{Z}_2=\{\overline{\mathpzc{X}},\overline{\mathpzc{I}}\}, \quad
\overline{\mathpzc{O}}=\sum_{i=1}^{N_{\mathrm{v}}} \overline{\sigma}^{y}_{i}/N_{\mathrm{v}}, 
\end{equation}
where $\overline{\mathpzc{X}}= \prod_{i}\overline{\sigma}^{x}_{i}, \overline{\mathpzc{I}}=\prod_{i}\overline{I}_{i}$. By employing the projection operator $\mathpzc{P}_{\mathrm{v}}$, these two quantities can be recast in the original basis as follows:   
\begin{eqnarray}
\overline{\mathpzc{O}}=\sum_{i=1}^{N_{\mathrm{v}}} &&\overline{\sigma}^{y}_{i}/N_{\mathrm{v}} \mapsto \mathpzc{O}=\sum_{\mathrm{v~links}}(\sigma_{i}^{x}\sigma_{j}^{y}+\sigma_{i}^{y}\sigma_{j}^{x})/2N_{\mathrm{v}},\nonumber \\
&&\overline{\mathpzc{X}}=\prod_{i} \overline{\sigma}^{x}_{i} \mapsto \mathpzc{X}=\prod_{\mathrm{v~links}}\sigma_{i}^{x}\sigma_{j}^{x},\nonumber \\
&&\overline{\mathpzc{I}}=\prod_{i} \overline{I_{i}} \mapsto \mathpzc{I}=\prod_{\mathrm{v~links}}I_{i}I_{j}. 
\label{X-symmetry}
\end{eqnarray}
The broken symmetry group $\mathbb{Z}_2=\{\mathpzc{X},\mathpzc{I}\}$ and the local order parameter $\mathpzc{O}$ uniquely characterize the $\mathfrak{B}$ phase in the sense that in this phase the symmetry $\mathpzc{X}$ is not preserved, and the local order parameter $\mathpzc{O}$ is nonzero. 

\begin{figure}
\includegraphics[width=0.80 \linewidth ]{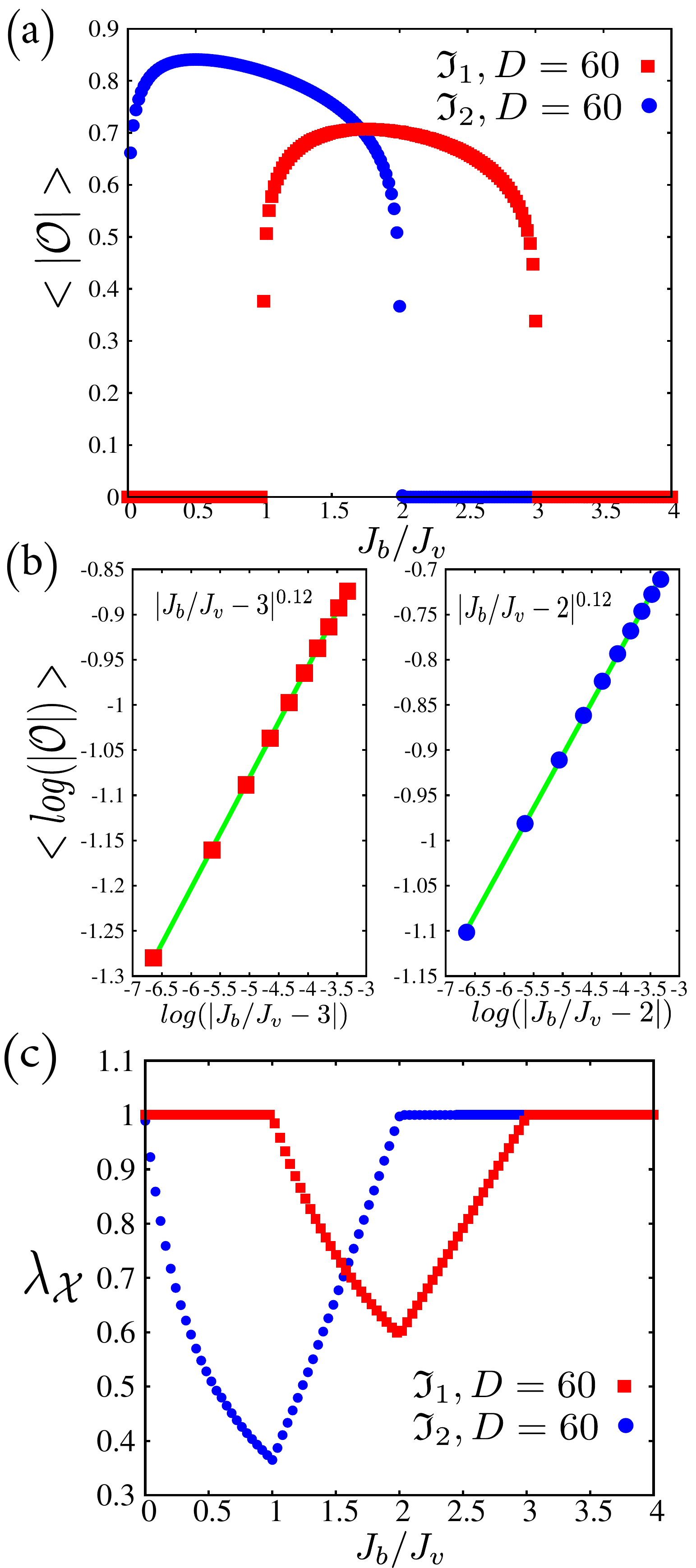}
\caption{(Color online). Local order parameter, its scaling close to the critical points and the broken symmetry along the paths $\{\mathfrak{I}_{1}, \mathfrak{I}_{2}\}$. (a) The local order parameter $\mathpzc{O}$ becomes nonzero within the $\mathfrak{B}$ phase, which indicates spontaneous symmetry breaking. (b) Log-log plot of the local order parameter $\mathpzc{O}$ versus $J_{\mathrm{v}}/J_{\mathrm{b}}$ in the vicinity of the critical point, which gives the magnetization exponent $\beta=0.12\pm0.01$. (c) $\lambda_{\mathpzc{X}}$ shows qualitative behavior of the symmetry $\mathpzc{X}$; when $\lambda_{\mathpzc{X}}<1$ it implies that $\mathpzc{X}$ is broken. In the $\mathfrak{B}$ phase, as expected, the symmetry $\mathpzc{X}$ is spontaneously broken.}
 \label{fig:Order-Phase-B}
\end{figure}

On the other hand, the $\mathfrak{A}$ and $\mathfrak{C}$ phases represent nondegenerate ground states, which both respect all symmetries, including $\mathpzc{X}$. This yields that $\mathpzc{O}$ is always zero within the $\mathfrak{A}$ and $\mathfrak{C}$ phases,
\begin{align*}
 \langle  \mathpzc{O} \rangle=\langle \mathpzc{O}\mathpzc{G} &\rangle=-\langle\mathpzc{G} \mathpzc{O} \rangle=
-\langle \mathpzc{O}\rangle, \quad \mathpzc{G}=\prod_{\mathrm{v~links}}\sigma^{z}_{i}I_{j}, \\
&\Rightarrow \langle  \mathpzc{O}\rangle=-\langle \mathpzc{O}\rangle=0, 
\end{align*}
where the operator $\mathpzc{G}$ is one of the symmetries of the model. As a result, the phase diagram of the compass ladder can be classified by the local order parameter $\mathpzc{O}$.

We have numerically plotted the local order parameter $\mathpzc{O}$ through the paths $\{\mathfrak{I}_{1}, \mathfrak{I}_{2}\}$ in Fig.~\ref{fig:Order-Phase-B}-(a). The plot indicates that whenever the $\mathfrak{B}$ phase appears (in the range of $1<J_{\mathrm{b}}/J_{\mathrm{v}}<3$ and $0<J_{\mathrm{b}}/J_{\mathrm{v}}<2$, respectively, for paths $\mathfrak{I}_{1}$ and $\mathfrak{I}_{2}$) the local order parameter $\mathpzc{O}$ becomes nonzero. In addition, $\mathpzc{O}$ decays when it approaches the boundaries of 
the $\mathfrak{B}$ phase---the points $J_{\mathrm{b}}/J_{\mathrm{v}}\in\{1,3\}$ and $J_{\mathrm{b}}/J_{\mathrm{v}}\in\{0,2\}$, respectively, for the paths $\mathfrak{I}_{1}$ and $\mathfrak{I}_{2}$. As plotted in Fig.~\ref{fig:Order-Phase-B}-(b), $\mathpzc{O}$ vanishes as
\begin{align}
\mathpzc{O}&\sim |J_{\mathrm{b}}/J_{\mathrm{v}}-3|^{\beta}, \\
\mathpzc{O}&\sim |J_{\mathrm{b}}/J_{\mathrm{v}}-2|^{\beta}, 
\end{align} 
in the vicinity of the boundary points $3$ and $2$ for the paths $\{\mathfrak{I}_{1}, \mathfrak{I}_{2}\}$, respectively, where $\beta=0.12\pm0.01$. This implies that the exponent $\beta$ is $1/8$, and the quantum phase transition is of the second-order type. The same results have been obtained by using nonlocal string order parameters in Ref.~\onlinecite{Feng:2007}. 

The behavior of the symmetry $\mathpzc{X}$ can be explicitly investigated by calculating the maximum eigenvalue of $\mathpzc{X}$-transfer matrix (i.e., $\lambda_{\mathpzc{X}}$), as plotted along the paths $\{\mathfrak{I}_{1}, \mathfrak{I}_{2}\}$ in Fig.~\ref{fig:Order-Phase-B}-(c). Again, whenever the $\mathfrak{B}$ phase appears, $\lambda_{\mathpzc{X}}$ becomes $<1$, implying that the symmetry has been broken. This observation agrees with the effective Hamiltonian (\ref{Ham:Ising}).

\subsection{Symmetry fractionalization}
\label{sec:symmetryfractionalization}

The technique of symmetry fractionalization provides a method to uniquely distinguish different SPT phases. 
This technique for $1$D gapped systems is complete, and provides a set of unique labels assigned to 
each SPT phase. These labels are obtained by transformation of the iMPS representation under  the symmetries of system. To clarify how these symmetries result in unique labels, we shall discuss two examples: $\mathbb{Z}_2\times\mathbb{Z}_2$ and $\mathpzc{K}$ symmetries.

Assume that the on-site symmetries $\mathpzc{G}=\prod_{i}g_{i}$ and $\mathpzc{H}=\prod_{i}h_{i}$ commute; $g_i h_j=h_j g_i$, and $g_i^{2}=h_i^{2}=\openone$ (for all $i$ and $j$). These symmetries are isomorphic to the $\mathbb{Z}_2$ symmetry group in the form of $\{\mathpzc{H}, \mathpzc{I}\}$ and $\{\mathpzc{G}, \mathpzc{I}\}$. One can combine these $\mathbb{Z}_2$ symmetry groups and form a $\mathbb{Z}_2\times\mathbb{Z}_2$ group with elements $\{\mathpzc{G}, \mathpzc{H}, \mathpzc{G}\mathpzc{H}, \mathpzc{I}\}$. If $\mathbb{Z}_2\times\mathbb{Z}_2$ is respected by the iMPS, the maximum eigenvalue of $\mathpzc{G}$- and $\mathpzc{H}$-transfer matrices should be equal to one ($\lambda_{\mathpzc{G}}= \lambda_{\mathpzc{H}}=1$) and Eq.~(\ref{Eq:SymmetryPreservation}) should be satisfied for the elements of the symmetry group. Equation~(\ref{Eq:SymmetryPreservation}) yields   
\begin{align}
& gh \widetilde{\Gamma} = U^{\dagger}_g U^{\dagger}_{h} \widetilde{\Gamma} U_{h}U_g, \nonumber \quad hg \widetilde{\Gamma} = U^{\dagger}_{h} U^{\dagger}_g \widetilde{\Gamma} U_{g}U_{h},\nonumber \\
 &\Rightarrow U_gU_{h} = e^{i\Omega_{gh}} U_{h}U_g, \label{Eq:Projective} \\
&  gg \widetilde{\Gamma}=\widetilde{\Gamma} = U^{\dagger}_g U^{\dagger}_{g} \widetilde{\Gamma} U_{g}U_g, \nonumber \quad hh \widetilde{\Gamma}=\widetilde{\Gamma} = U^{\dagger}_{h} U^{\dagger}_h \widetilde{\Gamma} U_{h}U_{h},\nonumber \\
& \Rightarrow U_gU_{g} = e^{i\Omega_{g}}, \quad U_{h}U_h=e^{i\Omega_{h}}, \label{Eq:Z2Z2} 
\end{align}
where the phase factor $e^{i\Omega_{gh}}$ is used to classify SPT phases (note that for simplicity the summations and phase ($e^{i\theta}$) have been ignored). By Eqs.~(\ref{Eq:Projective}) and (\ref{Eq:Z2Z2}), $e^{i\Omega_{gh}}$ can only be $\pm 1$. This allows two different orders: the SPT phase with $e^{i\Omega_{gh}}=-1$ and the trivial phase with $e^{i\Omega_{gh}}=+1$. Throughout the SPT (trivial) phase, we have $e^{i\Omega_{gh}}=-1(+1)$; the sign changes only upon a quantum phase transition. 
The minus sign also reveals that the SPT phase is protected by $\mathbb{Z}_2\times\mathbb{Z}_2$ symmetry; i.e., any perturbation which respects the symmetry cannot destroy the SPT phase. The two signs also represent two inequivalent projective 
representations of the $\mathbb{Z}_2\times \mathbb{Z}_2$ symmetry---see also Refs.~\onlinecite{Haghshenas:2014, Langari:2015}. 

Based on this observation, the topological order parameter $\mathpzc{O}_{\mathbb{Z}_2\times \mathbb{Z}_2}$ is introduced as follows:
\begin{align*}
\mathpzc{O}_{\mathbb{Z}_2\times \mathbb{Z}_2}&=
\left\{
\begin{array}{cl}
0&;\ \ \ |\lambda_{\mathpzc{G}}|<1\: \text{or} \: |\lambda_{\mathpzc{H}}|<1\\
(1/D)\mathrm{Tr}[{U_gU_hU_g^{\dagger}U_h^{\dagger}}]&;\ \ \ |\lambda_{\mathpzc{G}}|=|\lambda_{\mathpzc{H}}|=1
\end{array}\right.. \nonumber
\end{align*}
This order parameter only takes values $\{0,1,-1\}$, from which the phase can be characterized. Specifically, the values $0$, $1$, and $-1$, respectively, denote the symmetry-breaking, topologically-trivial, and SPT phases---corresponding to the $\mathbb{Z}_2\times \mathbb{Z}_2$ symmetry.  

If the iMPS is symmetric under the complex conjugate symmetry $\mathpzc{K}$, $\lambda_{\mathpzc{K}}=1$ and Eq.~(\ref{Eq:SymmetryPreservation}) becomes
\begin{equation}
({\widetilde{\Gamma}^{(m_{i})}})^{\ast}=e^{i\theta_{\mathpzc{K}}}U_{\mathpzc{K}}^{\dagger}{\widetilde{\Gamma}^{(m_{i})}} U_\mathpzc{K},
\label{Eq:ComplexConjugate}
\end{equation}
where $e^{i\theta_{\mathpzc{K}}}$ is a phase. Taking complex conjugate of Eq.~(\ref{Eq:ComplexConjugate}) and iterating this equation twice gives
\begin{align*}
\Gamma^{\ast}=U_{\mathpzc{K}}^{\dagger}\Gamma U_\mathpzc{K} \rightsquigarrow \Gamma=&{{U^{\dagger}_{\mathpzc{K}}}}^{\ast} \Gamma^{\ast} U_\mathpzc{K}^{\ast} \rightsquigarrow \Gamma=({U_\mathpzc{K}^{\ast}U_\mathpzc{K}})^{\dagger}\Gamma (U_\mathpzc{K}^{\ast}U_\mathpzc{K}) \\
&U_\mathpzc{K}^{\ast}U_\mathpzc{K}=e^{i\Omega_{\mathpzc{K}}} \openone,
\end{align*} 
(for simplicity index $(m_{i})$ and the arbitrary phase $e^{i\theta_{\mathpzc{K}}}$ have been ignored). Since $U_{\mathpzc{K}}$ is unitary, the phase $e^{i\Omega_{\mathpzc{K}}}$ becomes $\pm1$. Each of these signs denote a separate order. Specifically, $e^{i\Omega_{\mathpzc{K}}}=-1$ indicates an SPT phase protected by $\mathpzc{K}$, whereas $e^{i\Omega_{\mathpzc{K}}}=1$ indicates a topologically-trivial phase. Similar to $\mathpzc{O}_{\mathbb{Z}_2\times \mathbb{Z}_2}$, one can define a topological order parameter ($\mathpzc{O}_{\mathpzc{K}}$) that detects topological properties of the SPT phase protected by $\mathpzc{K}$,  
\begin{align*}
\mathpzc{O}_{\mathpzc{K}}&=
\left\{
\begin{array}{cl}
0&;\ \ \ |\lambda_{\mathpzc{K}}|<1\\\
(1/D)\mathrm{Tr}[U_{\mathpzc{K}}U_\mathpzc{K}^{\ast}]&;\ \ \ |\lambda_{\mathpzc{K}}|=1
\end{array}\right.. \nonumber
\end{align*}

\subsection{Topological order parameter}
\label{sec:topologicalorder}

The $\mathfrak{A}$ and $\mathfrak{C}$ phases have SPT orders, as we showed by our degenerate perturbation analysis. In this section, we investigate the topological aspects of theses phases, namely: (i) there is a specific $\mathbb{Z}_2\times \mathbb{Z}_2$ symmetry which protects both phases, and (ii) the complex-conjugate symmetry protects only the $\mathfrak{A}$ phase, which indicates that the $\mathfrak{A}$ and $\mathfrak{C}$ phases belong to different classes of SPT phases.    

The $\mathfrak{A}$ phase is characterized by the cluster Hamiltonian (\ref{Ham:ClusterY}), and its ground state belongs to an SPT phase protected by the following $\mathbb{Z}_2\times \mathbb{Z}_2$ symmetry group\cite{Else:2012, son:2011} (written in the logical basis):
\begin{align}
&\mathbb{Z}_2\times \mathbb{Z}_2 =\{\widehat{\mathpzc{G}},\widehat{\mathpzc{H}},\widehat{\mathpzc{G}}\widehat{\mathpzc{H}},\widehat{\mathpzc{I}}\},\\
\widehat{\mathpzc{G}}&=\prod_{2i} \widehat{\sigma}^{z}_{2i}, \quad \widehat{\mathpzc{H}}=\prod_{2i+1} \widehat{\sigma}^{z}_{2i+1}.
\end{align}     
Rewriting this symmetry group in the original basis results in  
 \begin{equation}
\mathpzc{G}=\prod_{\mathrm{v~links}} \sigma^{z}_{i}I_{j}, \quad \mathpzc{H}=\prod_{\mathrm{v~links}} I_{i}\sigma^{z}_{j},
\label{Eq:Z2Z2PhaseA}
\end{equation}
and $\mathbb{Z}_2\times \mathbb{Z}_2=\{\mathpzc{G},\mathpzc{H},\mathpzc{G}\mathpzc{H},\mathpzc{I}\}$. Thus, the associated topological order parameter $\mathpzc{O}_{\mathbb{Z}_2\times \mathbb{Z}_2}$ should take the value $-1$ (which signals the existence of SPT phase) within the whole region of the $\mathfrak{A}$ phase. 

\begin{figure}
\includegraphics[width=1.00 \linewidth]{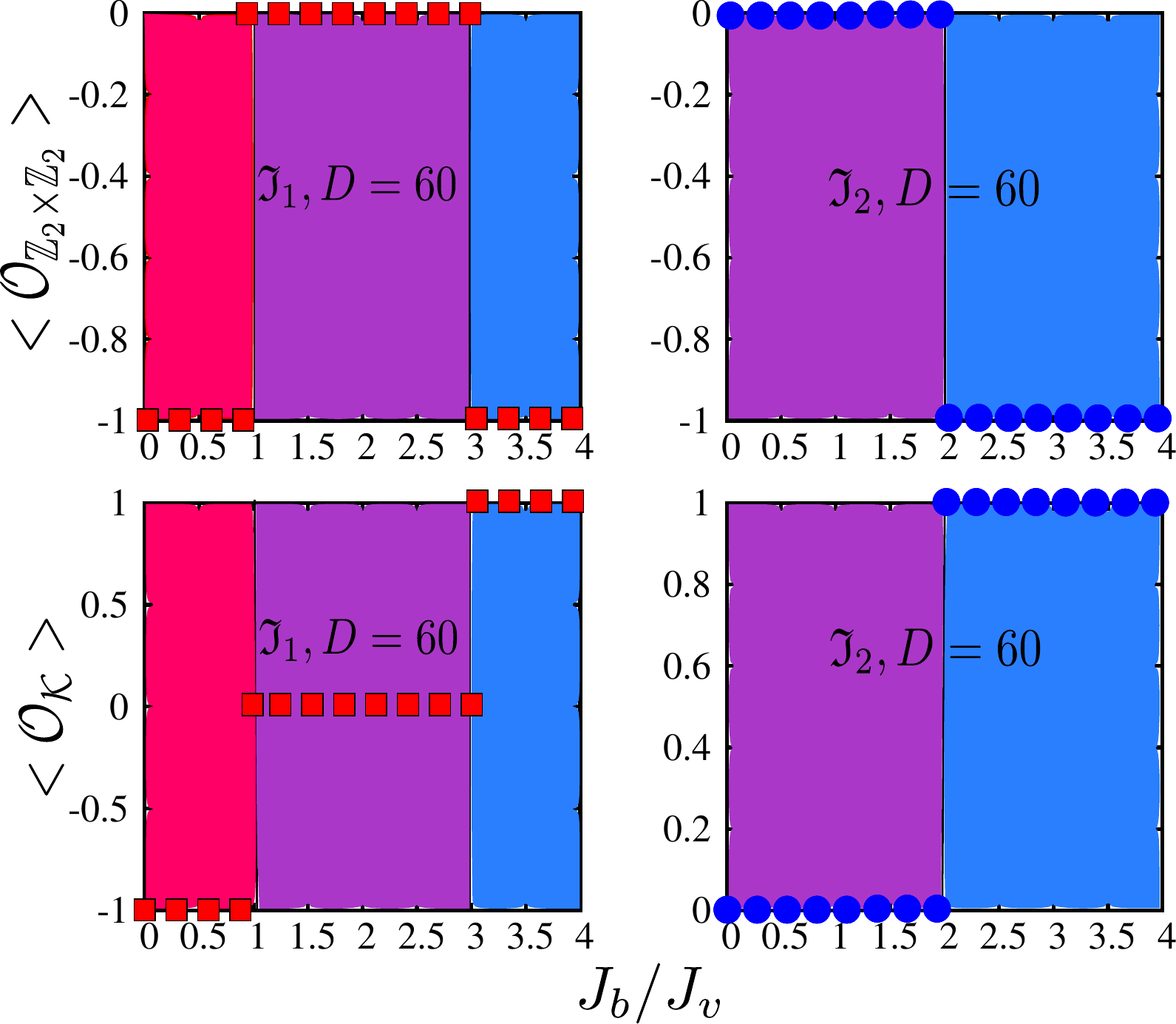}
\caption{(Color online). Topological order parameters $\mathpzc{O}_{\mathbb{Z}_2\times \mathbb{Z}_2}$ and $\mathpzc{O}_{\mathpzc{K}}$ for the paths $\{\mathfrak{I}_{1}, \mathfrak{I}_{2}\}$, (see the main text). The plot shows (i) both $\mathfrak{A}$ and $\mathfrak{C}$ phases are protected by the $\mathbb{Z}_2\times \mathbb{Z}_2$ symmetry, (ii) the complex-conjugate symmetry only protects the $\mathfrak{A}$ phase, and (iii) both symmetries are broken within the $\mathfrak{B}$ phase.}
\label{fig:phasefactor}
\end{figure}

It is straightforward to see that the $\mathbb{Z}_2\times \mathbb{Z}_2$ symmetry group of the $\mathfrak{C}$ phase has the exact form of Eq.~(\ref{Eq:Z2Z2PhaseA}). Thus, one concludes that $\mathpzc{O}_{\mathbb{Z}_2\times \mathbb{Z}_2}$ should be equal to $-1$ for both $\mathfrak{A}$ and $\mathfrak{C}$ phases, indicating the SPT phase protected by $\mathbb{Z}_2\times \mathbb{Z}_2$; and $0$ for the $\mathfrak{B}$ phase, implying the symmetry-breaking phase.

The topological order parameter $\mathpzc{O}_{\mathbb{Z}_2\times \mathbb{Z}_2}$ has been plotted in Fig.~\ref{fig:phasefactor} for the paths $\{\mathfrak{I}_{1}, \mathfrak{I}_{2}\}$. This plot confirms that $\mathpzc{O}_{\mathbb{Z}_2\times \mathbb{Z}_2}$ is $-1$ within the $\mathfrak{A}$ and $\mathfrak{C}$ phases, and $0$ within the $\mathfrak{B}$ phase. Note, however, that $\mathpzc{O}_{\mathbb{Z}_2\times \mathbb{Z}_2}$ does not distinguish the $\mathfrak{A}$ and $\mathfrak{C}$ phases; it only implies that both are of the SPT type. Thus we need to look for another topological order parameter to distinguish these phases.

The ground state of the cluster Hamiltonian (\ref{Ham:ClusterY}) has an exact iMPS form given by\cite{Garcia:2007}
\begin{equation}
\widetilde{\Gamma}^{(0)}=(1-i) \left(%
\begin{array}{cc}
  -1 & 1 \\
  1 & 1 \\
\end{array}%
\right)\;,\quad \widetilde{\Gamma}^{(1)}=(1+i)\left(%
\begin{array}{cc}
  1 & -1 \\
  1 & 1 \\
\end{array}%
\right)\;.
\end{equation}
Since this iMPS is symmetric under the complex-conjugate symmetry $\mathpzc{K}$, Eq.~(\ref{Eq:ComplexConjugate}) should hold. One can obtain (see Sec.~\ref{sec:iMPS}) that $U_{\mathpzc{K}}=\sigma^{y}$ and $e^{i\theta_{\mathpzc{K}}}=-i$. Moreover, $(\sigma^{y})({\sigma^{y}})^{\ast}=-\openone$, which immediately implies $\mathpzc{O}_{\mathpzc{K}}=-1$, demonstrating the SPT phase $\mathfrak{A}$ is protected by $\mathpzc{K}$. Nonetheless, we show that the $\mathfrak{C}$ phase is not protected by this symmetry.  

The iMPS form of the cluster Hamiltonian (\ref{Ham:ClusterX}) is expressed as follows:
\begin{equation}
\widetilde{\Gamma}^{(0)}=\left(%
\begin{array}{cc}
  -1 & 1 \\
  1 & 1 \\
\end{array}%
\right)\;,\quad \widetilde{\Gamma}^{(1)}=\left(%
\begin{array}{cc}
  1 & -1 \\
  1 & 1 \\
\end{array}%
\right)\;.
\end{equation}
This iMPS respects the complex-conjugate symmetry $\mathpzc{K}$, and Eq.~(\ref{Eq:ComplexConjugate}) is obviously satisfied by $U_{\mathpzc{K}}=\openone$ and $e^{i\theta_{\mathpzc{K}}}=1$. Hence, for this phase, $\mathpzc{O}_{\mathpzc{K}}=1$, implying that this phase is not protected by $\mathpzc{K}$. 

Summarizing, the topological order parameter $\mathpzc{O}_{\mathpzc{K}}$ is $-1$, $0$, and $1$ for the $\mathfrak{A}$, $\mathfrak{B}$, and $\mathfrak{C}$ phases, respectively (note that $\mathpzc{O}_{\mathpzc{K}}=0$ implies that $\mathpzc{K}$ symmetry has been broken). As depicted in Fig.~\ref{fig:phasefactor}, $\mathpzc{O}_{\mathpzc{K}}$ has been numerically 
calculated through the paths $\{\mathfrak{I}_{1}, \mathfrak{I}_{2}\}$. It demonstrates that the topological order parameter $\mathpzc{O}_{\mathpzc{K}}$ takes different values for each phase, thus it can truly (topologically) distinguish all three phases. This observation also indicates that one cannot adiabatically connect the $\mathfrak{C}$ and $\mathfrak{A}$ phases because they belong to different SPT classes. 

\section{1D Compass model}
\label{sec:compassmodel}

The ground space of the 1D compass model (Eq.~\ref{Eq:CompassModel}) for $J_{\mathrm{v}}=0$ can be represented as follows:
\begin{equation}
\mathpzc{V}_{b}=\{|\cdots\square_{i}\square_{i+1}\cdots\rangle ~|~
|\square_{i}\rangle=\alpha_{i}|  \nearrow\nearrow\rangle+\beta_{i}|\swarrow\swarrow\rangle \},
\end{equation}
where $|\square_{i}\rangle$ is defined on the $i$th blue link, and $\{\alpha_{i}, \beta_{i}\}$ are two arbitrary normalization factors  ($|\alpha_{i}|^{2}+|\beta_{i}|^{2}=1$). The ground-state subspace $\mathpzc{V}_{b}$, with degeneracy $|\mathpzc{V}_{b}|=2^{N/2}$, is stabilized by the symmetry $\mathpzc{X}$ (Eq.~\ref{X-symmetry}), that is, 
\begin{equation}
\mathpzc{X}|\Psi_{g}\rangle=|\Psi_{g}\rangle, \quad \forall|\Psi_{g}\rangle \in \mathpzc{V}_{b}.
\end{equation}
Thus, the symmetry group $\mathbb{Z}^{x}_{2}=\{\mathpzc{X}, \mathpzc{I}\}$ is respected at $J_{\mathrm{v}}=0$. Turning on the coupling parameter $J_{\mathrm{v}}$ does not affect this fact---that the symmetry $\mathpzc{X}$ stabilizes the ground space---up to the quantum phase transition point $J_{\mathrm{v}}=J_{\mathrm{b}}$ (see Fig.~\ref{fig:Kitaev phase}-(d)). This behavior is due to the fact that within blue-compass phase the gap is not closed. However, other symmetries of the compass model, 
such as $\mathpzc{Z}=\prod_{\mathrm{b~links}}\sigma_{i}^{z}\sigma_{j}^{z}$, time-reversal, and transnational invariance, are broken here. For instance, $\mathpzc{Z}$ does not stabilize the ground space $\mathpzc{V}_{b}$; rather, it rotates the elements of this space within itself,
\begin{align}
\mathpzc{Z}|\cdots\square_{i}\square_{i+1}\cdots\rangle&=|\cdots\square^{'}_{i}\square^{'}_{i+1}\cdots\rangle, \\
 |\square^{'}_{i}\rangle=\beta_{i}|\nearrow&\nearrow\rangle+\alpha_{i}|\swarrow\swarrow\rangle.
\end{align}
That is, the symmetry group $\mathbb{Z}^{z}_{2}=\{\mathpzc{Z}, \mathpzc{I}\}$ has been broken at $J_{\mathrm{v}}=0$. Similarly, because of the nonvanishing gap, this property is expected to hold within the blue-compass phase (see Fig.~\ref{fig:Kitaev phase}-(d)).

\begin{figure}
\includegraphics[width=6.5cm,height=13.5cm]{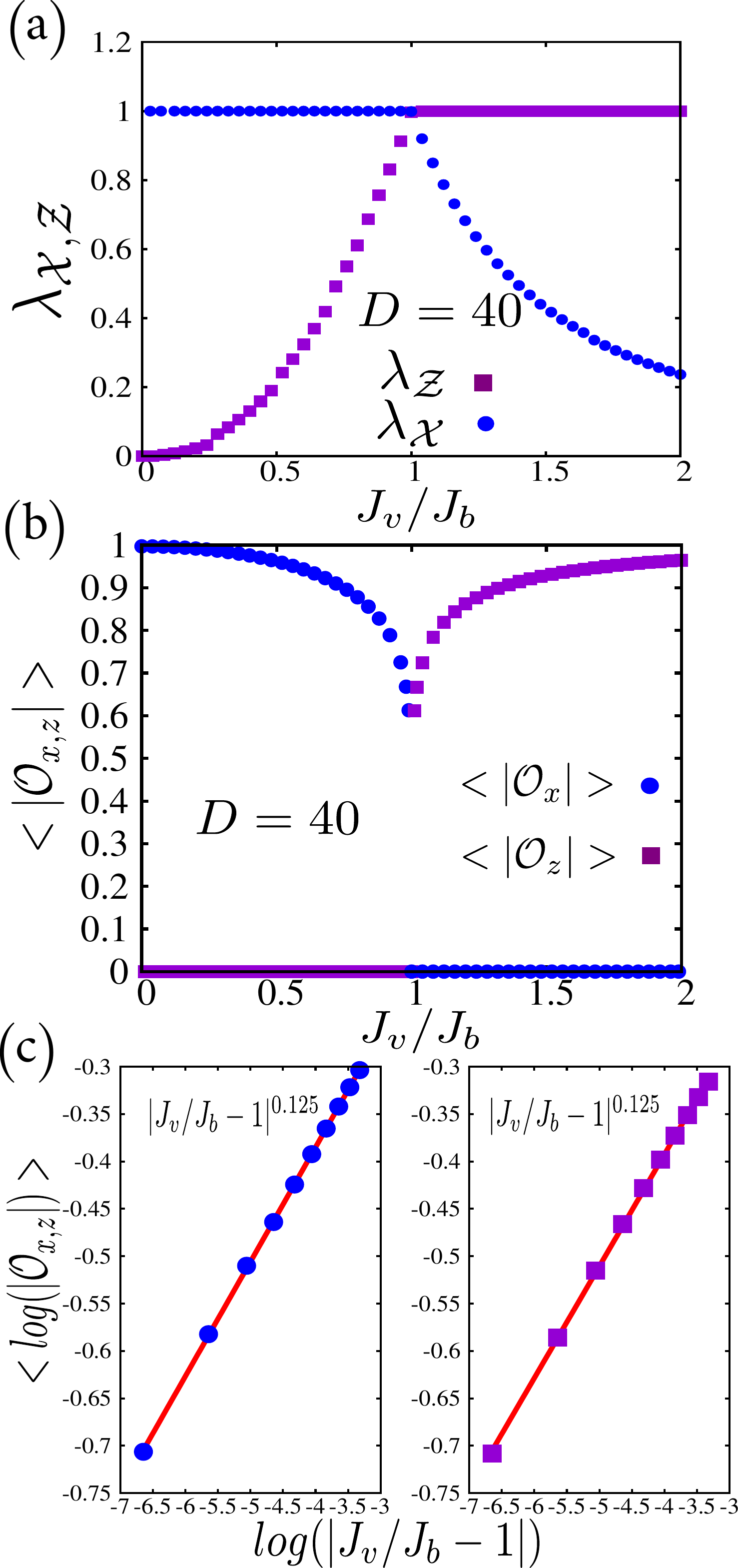} 
\caption{(Color online). Numerical analysis of the symmetries $\{\mathpzc{X}, \mathpzc{Z}\}$, local order parameters $\{\mathpzc{O}_{z}, \mathpzc{O}_{x}\}$, and their scaling. (a) Parameters $\{\lambda_{\mathpzc{X}}$ and $\lambda_{\mathpzc{Z}}\}$ show qualitative behavior of the symmetries $\{\mathpzc{X}$ and $\mathpzc{Z}\}$, respectively. Whenever these parameters become less than one, the associated symmetry is broken, and when they are equal to one, it means the associated symmetry is preserved. (b) The local order parameters $\{\mathpzc{O}_{z}$ and $\mathpzc{O}_{x}\}$ vs $J_{\mathrm{v}}/J_{\mathrm{b}}$. (c) Scaling of the order parameters near the critical point $J_{\mathrm{v}}/J_{\mathrm{b}}=1$. These scalings indicates that the magnetization exponent $\beta$ for both order parameters is $0.125\pm0.001$}
\label{fig:Compass}
\end{figure}

On the other hand, the symmetry $\mathpzc{Z}$ is stabilized by the ground space of the compass model at $J_{\mathrm{b}}=0$. This ground space is given by
\begin{equation}
\mathpzc{V}_{v}=\{|\cdots\blacksquare_{i}\blacksquare_{i+1}\cdots\rangle ~| ~|\blacksquare\rangle=\alpha'_{i}|  \uparrow\uparrow\rangle+\beta'_{i}|\downarrow\downarrow\rangle  \},
\end{equation}
where $|\blacksquare\rangle_{i}$ is defined on the $i$th violet link, and $\{\alpha'_{i}, \beta'_{i}\}$ are two arbitrary normalization coefficients. For nonzero values of $J_{\mathrm{b}}$, the symmetry group $\mathbb{Z}^{z}_{2}$ is respected throughout the violet-compass phase as long as the gap is nonzero. In addition, the symmetry group $\mathbb{Z}^{x}_{2}$ rotates 
elements of the ground space $\mathpzc{V}_{v}$, it becomes broken within the violet-compass phase. Thus the symmetry $\mathbb{Z}^{x}_{2}$ ($\mathbb{Z}^{z}_{2}$) is preserved within the blue-compass (violet-compass) phase, and is broken within violet-compass phase (blue-compass phase). 

To verify this observation, the parameters $\{\lambda_{\mathpzc{X}}, \lambda_{\mathpzc{Z}}\}$ have also been numerically calculated using the iTEBD method, which are shown in Fig.~\ref{fig:Compass}-(a). As expected, $\lambda_{\mathpzc{X}}$ ($\lambda_{\mathpzc{Z}}$) is equal to one throughout the blue (violet) phase, and is less than one otherwise. The breakdown of the aforementioned symmetry upon quantum phase transition can be captured by a local order parameter. Within the blue phase, where $\mathpzc{X}$ is preserved, the local order parameter $\mathpzc{O}_{z}=(1/N)\sum_{i} \sigma^{z}_{i}$ vanishes,
\begin{align*}
\langle  \mathpzc{O}_{z} \rangle=\langle &\mathpzc{O}_{z}\mathpzc{X} \rangle=-\langle\mathpzc{X} \mathpzc{O}_{z} \rangle=
-\langle \mathpzc{O}_{z}\rangle \\
&\Rightarrow \langle  \mathpzc{O}_{z}\rangle=-\langle \mathpzc{O}_{z}\rangle=0.
\end{align*} 
Whereas, the local order parameter $\mathpzc{O}_{z}$ becomes nonzero within the violet phase, which signals the breakdown of $\mathpzc{X}$ symmetry. As an example, for the ground state $|\cdots\uparrow\uparrow\cdots\rangle$ at 
$J_{\mathrm{b}}=0$, $\mathpzc{O}_{z}=1$. Hence, $\mathpzc{O}_{z}$ is a proper local order parameter to represent the 
breakdown of $\mathpzc{X}$ symmetry. In the same manner, one can show that the local order parameter $\mathpzc{O}_{x}=\frac{1}{N}\sum_{i} \sigma^{x}_{i}$ vanishes within the violet-compass phase (which comes from the preservation of 
the symmetry $\mathpzc{Z}$), and becomes nonzero within the blue-compass phase. We have numerically calculated the order parameters $\{\mathpzc{O}_{z}$ and $\mathpzc{O}_{x}\}$ by employing the iTEBD technique. As shown in Fig.~\ref{fig:Compass}-(b), the local order parameters $\{\mathpzc{O}_{z}, \mathpzc{O}_{x}\}$ behave exactly as explained before: $\mathpzc{O}_{x}$ vanishes throughout the violet-compass phase due to the preservation of the $\mathpzc{Z}$ symmetry, while it becomes nonzero in the blue-compass phase as a result of $\mathpzc{Z}$ symmetry breaking. Similar behavior is also observed for the order parameter $\mathpzc{O}_{z}$.

Both local order parameters $\{\mathpzc{O}_{z}$ and $\mathpzc{O}_{x}\}$ vanish as the quantum critical point is approached $J_{\mathrm{v}}/J_{\mathrm{b}}=1$,
\begin{align}
\mathpzc{O}_{x} &\sim |J_{\mathrm{v}}/J_{\mathrm{b}}-1|^{\beta'}, \\
\mathpzc{O}_{z}& \sim |J_{\mathrm{v}}/J_{\mathrm{b}}-1|^{\beta'},
\end{align} 
as shown in Fig.~\ref{fig:Compass}-(c), where $\beta'=0.125\pm0.001$. It justifies that the corresponding order parameter exponent is $\beta'=1/8$, associated to the a second-order quantum critical point. Thus the local order parameters $\mathpzc{O}_{z}$ and $\mathpzc{O}_{x}$ faithfully capture the relevant physics of the 1D compass model. These results are in agreement with Refs.~\onlinecite{Feng:2007, Wang:2015}, where physical quantities and phase characterization have been obtained by employing nonlocal string order parameters.      

\section{Summary and conclusions}
\label{sec:conclusion}

In this paper the topological classification of the phases and the associated quantum phase transitions in the compass ladder and $1$D compass models have been presented by employing degenerate perturbation theory, symmetry fractionalization, and numerical investigation. For each phase of the model (denoted by $\mathfrak{A}$, $\mathfrak{B}$, and $\mathfrak{C}$), we have derived an effective Hamiltonian based on degenerate perturbation theory. The $\mathfrak{A}$ and $\mathfrak{C}$ phases  
have been shown to be described by two cluster Hamiltonians written in different bases, whereas the $\mathfrak{B}$ phase has been shown to be represented by an Ising Hamiltonian. The cluster phase (specified by the ground state of the cluster model) is an SPT phase protected by a specific $\mathbb{Z}_2\times \mathbb{Z}_2$ symmetry, whereby we have assigned a set of labels to specify them. In other words, the set of unique labels of cluster phases have proven to be similar to that of the $\mathfrak{A}$ and $\mathfrak{C}$ phases. However, the $\mathfrak{A}$ and $\mathfrak{C}$ phases do not belong to the same class of an SPT phase: one of the phases is protected by the complex-conjugate symmetry, while the other is not. This observation has been verified by both numerical computations and analytical calculations.

We have shown that the $\mathfrak{B}$ phase is of topologically-trivial $\mathbb{Z}_2$-symmetry breaking type, characterized by a local order parameter. Having determined the form of the local order parameter and broken symmetry, we have concluded that (i) the phase diagram of model is characterized by a local order parameter, (ii) the quantum phase transition is associated to a spontaneous symmetry breaking (not topological), and (iii) the class of the quantum phase transition is in the Ising universality class, where the magnetization exponent is equal to $1/8$ and its type is of second order. We have also verified these observations numerically.

In addition, we have shown that the quantum phase transition in the $1$D compass model 
(a limiting case of the compass ladder) is accompanied by a partial symmetry breaking. 
Each of the phases of the $1$D compass model (the blue- and violet-compass phases) 
have been shown to respect part of a symmetry group; the blue- and violet-compass 
phases respect two different 
$\mathbb{Z}_{2}$-symmetry groups and break the other symmetries. 
Hence, upon the quantum phase transition, one of the $\mathbb{Z}_{2}$ symmetries 
is broken and other one is preserved. This partial symmetry breaking has been 
captured by local order parameters. By using numerical computations, 
we have shown that these local order parameters truly capture the 
quantum phase transition (as well as partial symmetry breaking) 
and its universality class (i.e., $\beta=1/8$). 
It is worth mentioning that this type of quantum phase transition is different from, 
e.g., transverse field Ising model, in which the whole symmetry group is being broken 
and system goes from a disordered phase to an ordered one. 

Our phase characterization of the compass ladder model has also revealed the nature (topological classes) of a number of the phases of the `Kitaev model on arbitrary-row brick wall lattice'---which is similar to that of the compass ladder model. Although the phase diagram of this model has been known, the nature of remaining phases and their corresponding quantum phase transitions are still largely unknown. An analysis based on our approach, especially symmetry fractionalization and degenerate perturbation theory, may shed some light on this direction. 

\begin{acknowledgments} 
This work was supported in part by Sharif University of Technology's Office of Vice President for Research.
\end{acknowledgments}

 \bibliography{refs-2.bib}

\begin{thebibliography}{48}%
\makeatletter
\providecommand \@ifxundefined [1]{%
 \@ifx{#1\undefined}
}%
\providecommand \@ifnum [1]{%
 \ifnum #1\expandafter \@firstoftwo
 \else \expandafter \@secondoftwo
 \fi
}%
\providecommand \@ifx [1]{%
 \ifx #1\expandafter \@firstoftwo
 \else \expandafter \@secondoftwo
 \fi
}%
\providecommand \natexlab [1]{#1}%
\providecommand \enquote  [1]{``#1''}%
\providecommand \bibnamefont  [1]{#1}%
\providecommand \bibfnamefont [1]{#1}%
\providecommand \citenamefont [1]{#1}%
\providecommand \href@noop [0]{\@secondoftwo}%
\providecommand \href [0]{\begingroup \@sanitize@url \@href}%
\providecommand \@href[1]{\@@startlink{#1}\@@href}%
\providecommand \@@href[1]{\endgroup#1\@@endlink}%
\providecommand \@sanitize@url [0]{\catcode `\\12\catcode `\$12\catcode
  `\&12\catcode `\#12\catcode `\^12\catcode `\_12\catcode `\%12\relax}%
\providecommand \@@startlink[1]{}%
\providecommand \@@endlink[0]{}%
\providecommand \url  [0]{\begingroup\@sanitize@url \@url }%
\providecommand \@url [1]{\endgroup\@href {#1}{\urlprefix }}%
\providecommand \urlprefix  [0]{URL }%
\providecommand \Eprint [0]{\href }%
\providecommand \doibase [0]{http://dx.doi.org/}%
\providecommand \selectlanguage [0]{\@gobble}%
\providecommand \bibinfo  [0]{\@secondoftwo}%
\providecommand \bibfield  [0]{\@secondoftwo}%
\providecommand \translation [1]{[#1]}%
\providecommand \BibitemOpen [0]{}%
\providecommand \bibitemStop [0]{}%
\providecommand \bibitemNoStop [0]{.\EOS\space}%
\providecommand \EOS [0]{\spacefactor3000\relax}%
\providecommand \BibitemShut  [1]{\csname bibitem#1\endcsname}%
\let\auto@bib@innerbib\@empty
\bibitem [{\citenamefont {Goldenfeld}(1992)}]{book:Goldenfeld}%
  \BibitemOpen
  \bibfield  {author} {\bibinfo {author} {\bibfnamefont {N.}~\bibnamefont
  {Goldenfeld}},\ }\href@noop {} {\emph {\bibinfo {title} {Lectures on Phase
  Transitions and the Renormalization Group}}}\ (\bibinfo  {publisher} {Perseus
  Books},\ \bibinfo {address} {Reading, MA},\ \bibinfo {year}
  {1992})\BibitemShut {NoStop}%
\bibitem [{\citenamefont {Tsui}\ \emph {et~al.}(1982)\citenamefont {Tsui},
  \citenamefont {Stormer},\ and\ \citenamefont {Gossard}}]{Tsui:1982}%
  \BibitemOpen
  \bibfield  {author} {\bibinfo {author} {\bibfnamefont {D.~C.}\ \bibnamefont
  {Tsui}}, \bibinfo {author} {\bibfnamefont {H.~L.}\ \bibnamefont {Stormer}}, \
  and\ \bibinfo {author} {\bibfnamefont {A.~C.}\ \bibnamefont {Gossard}},\
  }\href {\doibase 10.1103/PhysRevLett.48.1559} {\bibfield  {journal} {\bibinfo
   {journal} {Phys. Rev. Lett.}\ }\textbf {\bibinfo {volume} {48}},\ \bibinfo
  {pages} {1559} (\bibinfo {year} {1982})}\BibitemShut {NoStop}%
\bibitem [{\citenamefont {Kane}\ and\ \citenamefont {Mele}(2005)}]{Kane:2005}%
  \BibitemOpen
  \bibfield  {author} {\bibinfo {author} {\bibfnamefont {C.~L.}\ \bibnamefont
  {Kane}}\ and\ \bibinfo {author} {\bibfnamefont {E.~J.}\ \bibnamefont
  {Mele}},\ }\href {\doibase 10.1103/PhysRevLett.95.226801} {\bibfield
  {journal} {\bibinfo  {journal} {Phys. Rev. Lett.}\ }\textbf {\bibinfo
  {volume} {95}},\ \bibinfo {pages} {226801} (\bibinfo {year}
  {2005})}\BibitemShut {NoStop}%
\bibitem [{\citenamefont {Read}\ and\ \citenamefont
  {Sachdev}(1991)}]{Read:1991}%
  \BibitemOpen
  \bibfield  {author} {\bibinfo {author} {\bibfnamefont {N.}~\bibnamefont
  {Read}}\ and\ \bibinfo {author} {\bibfnamefont {S.}~\bibnamefont {Sachdev}},\
  }\href {\doibase 10.1103/PhysRevLett.66.1773} {\bibfield  {journal} {\bibinfo
   {journal} {Phys. Rev. Lett.}\ }\textbf {\bibinfo {volume} {66}},\ \bibinfo
  {pages} {1773} (\bibinfo {year} {1991})}\BibitemShut {NoStop}%
\bibitem [{\citenamefont {Kitaev}(2003)}]{kitaev:2003}%
  \BibitemOpen
  \bibfield  {author} {\bibinfo {author} {\bibfnamefont {A.}~\bibnamefont
  {Kitaev}},\ }\href {\doibase http://dx.doi.org/10.1016/S0003-4916(02)00018-0}
  {\bibfield  {journal} {\bibinfo  {journal} {Ann. Phys.}\ }\textbf {\bibinfo
  {volume} {303}},\ \bibinfo {pages} {2 } (\bibinfo {year} {2003})}\BibitemShut
  {NoStop}%
\bibitem [{\citenamefont {Nayak}\ \emph {et~al.}(2008)\citenamefont {Nayak},
  \citenamefont {Simon}, \citenamefont {Stern}, \citenamefont {Freedman},\ and\
  \citenamefont {Das~Sarma}}]{Nayak:2008}%
  \BibitemOpen
  \bibfield  {author} {\bibinfo {author} {\bibfnamefont {C.}~\bibnamefont
  {Nayak}}, \bibinfo {author} {\bibfnamefont {S.~H.}\ \bibnamefont {Simon}},
  \bibinfo {author} {\bibfnamefont {A.}~\bibnamefont {Stern}}, \bibinfo
  {author} {\bibfnamefont {M.}~\bibnamefont {Freedman}}, \ and\ \bibinfo
  {author} {\bibfnamefont {S.}~\bibnamefont {Das~Sarma}},\ }\href {\doibase
  10.1103/RevModPhys.80.1083} {\bibfield  {journal} {\bibinfo  {journal} {Rev.
  Mod. Phys.}\ }\textbf {\bibinfo {volume} {80}},\ \bibinfo {pages} {1083}
  (\bibinfo {year} {2008})}\BibitemShut {NoStop}%
\bibitem [{\citenamefont {Wen}\ and\ \citenamefont {Niu}(1990)}]{Wen:1990}%
  \BibitemOpen
  \bibfield  {author} {\bibinfo {author} {\bibfnamefont {X.~G.}\ \bibnamefont
  {Wen}}\ and\ \bibinfo {author} {\bibfnamefont {Q.}~\bibnamefont {Niu}},\
  }\href {\doibase 10.1103/PhysRevB.41.9377} {\bibfield  {journal} {\bibinfo
  {journal} {Phys. Rev. B}\ }\textbf {\bibinfo {volume} {41}},\ \bibinfo
  {pages} {9377} (\bibinfo {year} {1990})}\BibitemShut {NoStop}%
\bibitem [{\citenamefont {Wen}(2007)}]{Wen:2007}%
  \BibitemOpen
  \bibfield  {author} {\bibinfo {author} {\bibfnamefont {X.-G.}\ \bibnamefont
  {Wen}},\ }\href@noop {} {\emph {\bibinfo {title} {{Quantum Field Theory of
  Many-body Systems: From the Origin of Sound to an Origin of Light and
  Electrons}}}}\ (\bibinfo  {publisher} {Oxford University Press},\ \bibinfo
  {address} {Oxford},\ \bibinfo {year} {2007})\BibitemShut {NoStop}%
\bibitem [{\citenamefont {Chen}\ \emph {et~al.}(2010)\citenamefont {Chen},
  \citenamefont {Gu},\ and\ \citenamefont {Wen}}]{Chen:2010}%
  \BibitemOpen
  \bibfield  {author} {\bibinfo {author} {\bibfnamefont {X.}~\bibnamefont
  {Chen}}, \bibinfo {author} {\bibfnamefont {Z.-C.}\ \bibnamefont {Gu}}, \ and\
  \bibinfo {author} {\bibfnamefont {X.-G.}\ \bibnamefont {Wen}},\ }\href
  {\doibase 10.1103/PhysRevB.82.155138} {\bibfield  {journal} {\bibinfo
  {journal} {Phys. Rev. B}\ }\textbf {\bibinfo {volume} {82}},\ \bibinfo
  {pages} {155138} (\bibinfo {year} {2010})}\BibitemShut {NoStop}%
\bibitem [{\citenamefont {Chen}\ \emph
  {et~al.}(2011{\natexlab{a}})\citenamefont {Chen}, \citenamefont {Gu},\ and\
  \citenamefont {Wen}}]{Chen:2011:Jan}%
  \BibitemOpen
  \bibfield  {author} {\bibinfo {author} {\bibfnamefont {X.}~\bibnamefont
  {Chen}}, \bibinfo {author} {\bibfnamefont {Z.-C.}\ \bibnamefont {Gu}}, \ and\
  \bibinfo {author} {\bibfnamefont {X.-G.}\ \bibnamefont {Wen}},\ }\href
  {\doibase 10.1103/PhysRevB.83.035107} {\bibfield  {journal} {\bibinfo
  {journal} {Phys. Rev. B}\ }\textbf {\bibinfo {volume} {83}},\ \bibinfo
  {pages} {035107} (\bibinfo {year} {2011}{\natexlab{a}})}\BibitemShut
  {NoStop}%
\bibitem [{\citenamefont {Chen}\ \emph
  {et~al.}(2011{\natexlab{b}})\citenamefont {Chen}, \citenamefont {Gu},\ and\
  \citenamefont {Wen}}]{Chen:2011}%
  \BibitemOpen
  \bibfield  {author} {\bibinfo {author} {\bibfnamefont {X.}~\bibnamefont
  {Chen}}, \bibinfo {author} {\bibfnamefont {Z.-C.}\ \bibnamefont {Gu}}, \ and\
  \bibinfo {author} {\bibfnamefont {X.-G.}\ \bibnamefont {Wen}},\ }\href
  {\doibase 10.1103/PhysRevB.84.235128} {\bibfield  {journal} {\bibinfo
  {journal} {Phys. Rev. B}\ }\textbf {\bibinfo {volume} {84}},\ \bibinfo
  {pages} {235128} (\bibinfo {year} {2011}{\natexlab{b}})}\BibitemShut
  {NoStop}%
\bibitem [{\citenamefont {Schuch}\ \emph {et~al.}(2011)\citenamefont {Schuch},
  \citenamefont {Perez-Garcia},\ and\ \citenamefont {Cirac}}]{Schuch:2011}%
  \BibitemOpen
  \bibfield  {author} {\bibinfo {author} {\bibfnamefont {N.}~\bibnamefont
  {Schuch}}, \bibinfo {author} {\bibfnamefont {D.}~\bibnamefont
  {Perez-Garcia}}, \ and\ \bibinfo {author} {\bibfnamefont {I.}~\bibnamefont
  {Cirac}},\ }\href {\doibase 10.1103/PhysRevB.84.165139} {\bibfield  {journal}
  {\bibinfo  {journal} {Phys. Rev. B}\ }\textbf {\bibinfo {volume} {84}},\
  \bibinfo {pages} {165139} (\bibinfo {year} {2011})}\BibitemShut {NoStop}%
\bibitem [{\citenamefont {Pollmann}\ \emph {et~al.}(2010)\citenamefont
  {Pollmann}, \citenamefont {Turner}, \citenamefont {Berg},\ and\ \citenamefont
  {Oshikawa}}]{Pollmann:2010}%
  \BibitemOpen
  \bibfield  {author} {\bibinfo {author} {\bibfnamefont {F.}~\bibnamefont
  {Pollmann}}, \bibinfo {author} {\bibfnamefont {A.~M.}\ \bibnamefont
  {Turner}}, \bibinfo {author} {\bibfnamefont {E.}~\bibnamefont {Berg}}, \ and\
  \bibinfo {author} {\bibfnamefont {M.}~\bibnamefont {Oshikawa}},\ }\href
  {\doibase 10.1103/PhysRevB.81.064439} {\bibfield  {journal} {\bibinfo
  {journal} {Phys. Rev. B}\ }\textbf {\bibinfo {volume} {81}},\ \bibinfo
  {pages} {064439} (\bibinfo {year} {2010})}\BibitemShut {NoStop}%
\bibitem [{\citenamefont {Vidal}(2007)}]{Vidal:2006}%
  \BibitemOpen
  \bibfield  {author} {\bibinfo {author} {\bibfnamefont {G.}~\bibnamefont
  {Vidal}},\ }\href {\doibase 10.1103/PhysRevLett.98.070201} {\bibfield
  {journal} {\bibinfo  {journal} {Phys. Rev. Lett.}\ }\textbf {\bibinfo
  {volume} {98}},\ \bibinfo {pages} {070201} (\bibinfo {year}
  {2007})}\BibitemShut {NoStop}%
\bibitem [{\citenamefont {Schollw{\"o}ck}(2011)}]{Schollwock:2011}%
  \BibitemOpen
  \bibfield  {author} {\bibinfo {author} {\bibfnamefont {U.}~\bibnamefont
  {Schollw{\"o}ck}},\ }\href
  {http://www.sciencedirect.com/science/article/pii/S0003491610001752}
  {\bibfield  {journal} {\bibinfo  {journal} {Ann. Phys.}\ }\textbf {\bibinfo
  {volume} {326}},\ \bibinfo {pages} {96} (\bibinfo {year} {2011})}\BibitemShut
  {NoStop}%
\bibitem [{\citenamefont {Haegeman}\ \emph {et~al.}(2012)\citenamefont
  {Haegeman}, \citenamefont {Perez-Garcia}, \citenamefont {Cirac},\ and\
  \citenamefont {Schuch}}]{Haegeman:2012}%
  \BibitemOpen
  \bibfield  {author} {\bibinfo {author} {\bibfnamefont {J.}~\bibnamefont
  {Haegeman}}, \bibinfo {author} {\bibfnamefont {D.}~\bibnamefont
  {Perez-Garcia}}, \bibinfo {author} {\bibfnamefont {I.}~\bibnamefont {Cirac}},
  \ and\ \bibinfo {author} {\bibfnamefont {N.}~\bibnamefont {Schuch}},\ }\href
  {\doibase 10.1103/PhysRevLett.109.050402} {\bibfield  {journal} {\bibinfo
  {journal} {Phys. Rev. Lett.}\ }\textbf {\bibinfo {volume} {109}},\ \bibinfo
  {pages} {050402} (\bibinfo {year} {2012})}\BibitemShut {NoStop}%
\bibitem [{\citenamefont {Pollmann}\ and\ \citenamefont
  {Turner}(2012)}]{Pollmann:2012}%
  \BibitemOpen
  \bibfield  {author} {\bibinfo {author} {\bibfnamefont {F.}~\bibnamefont
  {Pollmann}}\ and\ \bibinfo {author} {\bibfnamefont {A.~M.}\ \bibnamefont
  {Turner}},\ }\href {\doibase 10.1103/PhysRevB.86.125441} {\bibfield
  {journal} {\bibinfo  {journal} {Phys. Rev. B}\ }\textbf {\bibinfo {volume}
  {86}},\ \bibinfo {pages} {125441} (\bibinfo {year} {2012})}\BibitemShut
  {NoStop}%
\bibitem [{\citenamefont {Levin}\ and\ \citenamefont {Wen}(2005)}]{Levin:2005}%
  \BibitemOpen
  \bibfield  {author} {\bibinfo {author} {\bibfnamefont {M.~A.}\ \bibnamefont
  {Levin}}\ and\ \bibinfo {author} {\bibfnamefont {X.-G.}\ \bibnamefont
  {Wen}},\ }\href {\doibase 10.1103/PhysRevB.71.045110} {\bibfield  {journal}
  {\bibinfo  {journal} {Phys. Rev. B}\ }\textbf {\bibinfo {volume} {71}},\
  \bibinfo {pages} {045110} (\bibinfo {year} {2005})}\BibitemShut {NoStop}%
\bibitem [{\citenamefont {Lin}\ and\ \citenamefont {Levin}(2014)}]{Lin:2014}%
  \BibitemOpen
  \bibfield  {author} {\bibinfo {author} {\bibfnamefont {C.-H.}\ \bibnamefont
  {Lin}}\ and\ \bibinfo {author} {\bibfnamefont {M.}~\bibnamefont {Levin}},\
  }\href {\doibase 10.1103/PhysRevB.89.195130} {\bibfield  {journal} {\bibinfo
  {journal} {Phys. Rev. B}\ }\textbf {\bibinfo {volume} {89}},\ \bibinfo
  {pages} {195130} (\bibinfo {year} {2014})}\BibitemShut {NoStop}%
\bibitem [{\citenamefont {Fendley}\ and\ \citenamefont
  {Fradkin}(2005)}]{Fendley:2005}%
  \BibitemOpen
  \bibfield  {author} {\bibinfo {author} {\bibfnamefont {P.}~\bibnamefont
  {Fendley}}\ and\ \bibinfo {author} {\bibfnamefont {E.}~\bibnamefont
  {Fradkin}},\ }\href {\doibase 10.1103/PhysRevB.72.024412} {\bibfield
  {journal} {\bibinfo  {journal} {Phys. Rev. B}\ }\textbf {\bibinfo {volume}
  {72}},\ \bibinfo {pages} {024412} (\bibinfo {year} {2005})}\BibitemShut
  {NoStop}%
\bibitem [{\citenamefont {Kitaev}(2006)}]{Kitaev:2006:January}%
  \BibitemOpen
  \bibfield  {author} {\bibinfo {author} {\bibfnamefont {A.}~\bibnamefont
  {Kitaev}},\ }\href {\doibase http://dx.doi.org/10.1016/j.aop.2005.10.005}
  {\bibfield  {journal} {\bibinfo  {journal} {Ann. Phys.}\ }\textbf {\bibinfo
  {volume} {321}},\ \bibinfo {pages} {2 } (\bibinfo {year} {2006})}\BibitemShut
  {NoStop}%
\bibitem [{\citenamefont {Lee}\ \emph {et~al.}(2014)\citenamefont {Lee},
  \citenamefont {Schaffer}, \citenamefont {Bhattacharjee},\ and\ \citenamefont
  {Kim}}]{Lee:2014}%
  \BibitemOpen
  \bibfield  {author} {\bibinfo {author} {\bibfnamefont {E.~K.-H.}\
  \bibnamefont {Lee}}, \bibinfo {author} {\bibfnamefont {R.}~\bibnamefont
  {Schaffer}}, \bibinfo {author} {\bibfnamefont {S.}~\bibnamefont
  {Bhattacharjee}}, \ and\ \bibinfo {author} {\bibfnamefont {Y.~B.}\
  \bibnamefont {Kim}},\ }\href {\doibase 10.1103/PhysRevB.89.045117} {\bibfield
   {journal} {\bibinfo  {journal} {Phys. Rev. B}\ }\textbf {\bibinfo {volume}
  {89}},\ \bibinfo {pages} {045117} (\bibinfo {year} {2014})}\BibitemShut
  {NoStop}%
\bibitem [{\citenamefont {Chaloupka}\ \emph {et~al.}(2013)\citenamefont
  {Chaloupka}, \citenamefont {Checkrelse}, \citenamefont {Jackeli},\ and\
  \citenamefont {Khaliullin}}]{Chaloupka:2013}%
  \BibitemOpen
  \bibfield  {author} {\bibinfo {author} {\bibfnamefont {J.}~\bibnamefont
  {Chaloupka}}, \bibinfo {author} {\bibfnamefont {V.}~\bibnamefont
  {Checkrelse}}, \bibinfo {author} {\bibfnamefont {G.}~\bibnamefont {Jackeli}},
  \ and\ \bibinfo {author} {\bibfnamefont {G.}~\bibnamefont {Khaliullin}},\
  }\href {\doibase 10.1103/PhysRevLett.110.097204} {\bibfield  {journal}
  {\bibinfo  {journal} {Phys. Rev. Lett.}\ }\textbf {\bibinfo {volume} {110}},\
  \bibinfo {pages} {097204} (\bibinfo {year} {2013})}\BibitemShut {NoStop}%
\bibitem [{\citenamefont {Reuther}\ \emph {et~al.}(2011)\citenamefont
  {Reuther}, \citenamefont {Thomale},\ and\ \citenamefont
  {Trebst}}]{Reuther:2011}%
  \BibitemOpen
  \bibfield  {author} {\bibinfo {author} {\bibfnamefont {J.}~\bibnamefont
  {Reuther}}, \bibinfo {author} {\bibfnamefont {R.}~\bibnamefont {Thomale}}, \
  and\ \bibinfo {author} {\bibfnamefont {S.}~\bibnamefont {Trebst}},\ }\href
  {\doibase 10.1103/PhysRevB.84.100406} {\bibfield  {journal} {\bibinfo
  {journal} {Phys. Rev. B}\ }\textbf {\bibinfo {volume} {84}},\ \bibinfo
  {pages} {100406} (\bibinfo {year} {2011})}\BibitemShut {NoStop}%
\bibitem [{\citenamefont {Osorio~Iregui}\ \emph {et~al.}(2014)\citenamefont
  {Osorio~Iregui}, \citenamefont {Corboz},\ and\ \citenamefont
  {Troyer}}]{Osorio:2014}%
  \BibitemOpen
  \bibfield  {author} {\bibinfo {author} {\bibfnamefont {J.}~\bibnamefont
  {Osorio~Iregui}}, \bibinfo {author} {\bibfnamefont {P.}~\bibnamefont
  {Corboz}}, \ and\ \bibinfo {author} {\bibfnamefont {M.}~\bibnamefont
  {Troyer}},\ }\href {\doibase 10.1103/PhysRevB.90.195102} {\bibfield
  {journal} {\bibinfo  {journal} {Phys. Rev. B}\ }\textbf {\bibinfo {volume}
  {90}},\ \bibinfo {pages} {195102} (\bibinfo {year} {2014})}\BibitemShut
  {NoStop}%
\bibitem [{\citenamefont {Barkeshli}\ \emph {et~al.}(2015)\citenamefont
  {Barkeshli}, \citenamefont {Jiang}, \citenamefont {Thomale},\ and\
  \citenamefont {Qi}}]{Barkeshli:2015}%
  \BibitemOpen
  \bibfield  {author} {\bibinfo {author} {\bibfnamefont {M.}~\bibnamefont
  {Barkeshli}}, \bibinfo {author} {\bibfnamefont {H.-C.}\ \bibnamefont
  {Jiang}}, \bibinfo {author} {\bibfnamefont {R.}~\bibnamefont {Thomale}}, \
  and\ \bibinfo {author} {\bibfnamefont {X.-L.}\ \bibnamefont {Qi}},\ }\href
  {\doibase 10.1103/PhysRevLett.114.026401} {\bibfield  {journal} {\bibinfo
  {journal} {Phys. Rev. Lett.}\ }\textbf {\bibinfo {volume} {114}},\ \bibinfo
  {pages} {026401} (\bibinfo {year} {2015})}\BibitemShut {NoStop}%
\bibitem [{\citenamefont {Feng}\ \emph {et~al.}(2007)\citenamefont {Feng},
  \citenamefont {Zhang},\ and\ \citenamefont {Xiang}}]{Feng:2007}%
  \BibitemOpen
  \bibfield  {author} {\bibinfo {author} {\bibfnamefont {X.-Y.}\ \bibnamefont
  {Feng}}, \bibinfo {author} {\bibfnamefont {G.-M.}\ \bibnamefont {Zhang}}, \
  and\ \bibinfo {author} {\bibfnamefont {T.}~\bibnamefont {Xiang}},\ }\href
  {\doibase 10.1103/PhysRevLett.98.087204} {\bibfield  {journal} {\bibinfo
  {journal} {Phys. Rev. Lett.}\ }\textbf {\bibinfo {volume} {98}},\ \bibinfo
  {pages} {087204} (\bibinfo {year} {2007})}\BibitemShut {NoStop}%
\bibitem [{\citenamefont {Brzezicki}\ \emph {et~al.}(2007)\citenamefont
  {Brzezicki}, \citenamefont {Dziarmaga},\ and\ \citenamefont
  {Ole\ifmmode~\acute{s}\else \'{s}\fi{}}}]{Brzezicki:2007}%
  \BibitemOpen
  \bibfield  {author} {\bibinfo {author} {\bibfnamefont {W.}~\bibnamefont
  {Brzezicki}}, \bibinfo {author} {\bibfnamefont {J.}~\bibnamefont
  {Dziarmaga}}, \ and\ \bibinfo {author} {\bibfnamefont {A.~M.}\ \bibnamefont
  {Ole\ifmmode~\acute{s}\else \'{s}\fi{}}},\ }\href {\doibase
  10.1103/PhysRevB.75.134415} {\bibfield  {journal} {\bibinfo  {journal} {Phys.
  Rev. B}\ }\textbf {\bibinfo {volume} {75}},\ \bibinfo {pages} {134415}
  (\bibinfo {year} {2007})}\BibitemShut {NoStop}%
\bibitem [{\citenamefont {Duan}\ \emph {et~al.}(2003)\citenamefont {Duan},
  \citenamefont {Demler},\ and\ \citenamefont {Lukin}}]{Duan:2003}%
  \BibitemOpen
  \bibfield  {author} {\bibinfo {author} {\bibfnamefont {L.-M.}\ \bibnamefont
  {Duan}}, \bibinfo {author} {\bibfnamefont {E.}~\bibnamefont {Demler}}, \ and\
  \bibinfo {author} {\bibfnamefont {M.~D.}\ \bibnamefont {Lukin}},\ }\href
  {\doibase 10.1103/PhysRevLett.91.090402} {\bibfield  {journal} {\bibinfo
  {journal} {Phys. Rev. Lett.}\ }\textbf {\bibinfo {volume} {91}},\ \bibinfo
  {pages} {090402} (\bibinfo {year} {2003})}\BibitemShut {NoStop}%
\bibitem [{\citenamefont {You}\ \emph {et~al.}(2010)\citenamefont {You},
  \citenamefont {Shi}, \citenamefont {Hu},\ and\ \citenamefont
  {Nori}}]{You:2010}%
  \BibitemOpen
  \bibfield  {author} {\bibinfo {author} {\bibfnamefont {J.~Q.}\ \bibnamefont
  {You}}, \bibinfo {author} {\bibfnamefont {X.-F.}\ \bibnamefont {Shi}},
  \bibinfo {author} {\bibfnamefont {X.}~\bibnamefont {Hu}}, \ and\ \bibinfo
  {author} {\bibfnamefont {F.}~\bibnamefont {Nori}},\ }\href {\doibase
  10.1103/PhysRevB.81.014505} {\bibfield  {journal} {\bibinfo  {journal} {Phys.
  Rev. B}\ }\textbf {\bibinfo {volume} {81}},\ \bibinfo {pages} {014505}
  (\bibinfo {year} {2010})}\BibitemShut {NoStop}%
\bibitem [{\citenamefont {Saket}\ \emph {et~al.}(2010)\citenamefont {Saket},
  \citenamefont {Hassan},\ and\ \citenamefont {Shankar}}]{Saket:2010}%
  \BibitemOpen
  \bibfield  {author} {\bibinfo {author} {\bibfnamefont {A.}~\bibnamefont
  {Saket}}, \bibinfo {author} {\bibfnamefont {S.~R.}\ \bibnamefont {Hassan}}, \
  and\ \bibinfo {author} {\bibfnamefont {R.}~\bibnamefont {Shankar}},\ }\href
  {\doibase 10.1103/PhysRevB.82.174409} {\bibfield  {journal} {\bibinfo
  {journal} {Phys. Rev. B}\ }\textbf {\bibinfo {volume} {82}},\ \bibinfo
  {pages} {174409} (\bibinfo {year} {2010})}\BibitemShut {NoStop}%
\bibitem [{\citenamefont {Tserkovnyak}\ and\ \citenamefont
  {Loss}(2011)}]{Tserkovnyak:2011}%
  \BibitemOpen
  \bibfield  {author} {\bibinfo {author} {\bibfnamefont {Y.}~\bibnamefont
  {Tserkovnyak}}\ and\ \bibinfo {author} {\bibfnamefont {D.}~\bibnamefont
  {Loss}},\ }\href {\doibase 10.1103/PhysRevA.84.032333} {\bibfield  {journal}
  {\bibinfo  {journal} {Phys. Rev. A}\ }\textbf {\bibinfo {volume} {84}},\
  \bibinfo {pages} {032333} (\bibinfo {year} {2011})}\BibitemShut {NoStop}%
\bibitem [{\citenamefont {He}\ and\ \citenamefont {Chen}(2013)}]{He:2013}%
  \BibitemOpen
  \bibfield  {author} {\bibinfo {author} {\bibfnamefont {Y.-C.}\ \bibnamefont
  {He}}\ and\ \bibinfo {author} {\bibfnamefont {Y.}~\bibnamefont {Chen}},\
  }\href {\doibase 10.1103/PhysRevB.88.180402} {\bibfield  {journal} {\bibinfo
  {journal} {Phys. Rev. B}\ }\textbf {\bibinfo {volume} {88}},\ \bibinfo
  {pages} {180402} (\bibinfo {year} {2013})}\BibitemShut {NoStop}%
\bibitem [{\citenamefont {Pedrocchi}\ \emph {et~al.}(2012)\citenamefont
  {Pedrocchi}, \citenamefont {Chesi}, \citenamefont {Gangadharaiah},\ and\
  \citenamefont {Loss}}]{Pedrocchi:2012}%
  \BibitemOpen
  \bibfield  {author} {\bibinfo {author} {\bibfnamefont {F.~L.}\ \bibnamefont
  {Pedrocchi}}, \bibinfo {author} {\bibfnamefont {S.}~\bibnamefont {Chesi}},
  \bibinfo {author} {\bibfnamefont {S.}~\bibnamefont {Gangadharaiah}}, \ and\
  \bibinfo {author} {\bibfnamefont {D.}~\bibnamefont {Loss}},\ }\href {\doibase
  10.1103/PhysRevB.86.205412} {\bibfield  {journal} {\bibinfo  {journal} {Phys.
  Rev. B}\ }\textbf {\bibinfo {volume} {86}},\ \bibinfo {pages} {205412}
  (\bibinfo {year} {2012})}\BibitemShut {NoStop}%
\bibitem [{\citenamefont {Karimipour}(2009)}]{Karimipour:2009}%
  \BibitemOpen
  \bibfield  {author} {\bibinfo {author} {\bibfnamefont {V.}~\bibnamefont
  {Karimipour}},\ }\href {\doibase 10.1103/PhysRevB.79.214435} {\bibfield
  {journal} {\bibinfo  {journal} {Phys. Rev. B}\ }\textbf {\bibinfo {volume}
  {79}},\ \bibinfo {pages} {214435} (\bibinfo {year} {2009})}\BibitemShut
  {NoStop}%
\bibitem [{\citenamefont {Karimipour}\ \emph {et~al.}(2013)\citenamefont
  {Karimipour}, \citenamefont {Memarzadeh},\ and\ \citenamefont
  {Zarkeshian}}]{Karimipour:2013}%
  \BibitemOpen
  \bibfield  {author} {\bibinfo {author} {\bibfnamefont {V.}~\bibnamefont
  {Karimipour}}, \bibinfo {author} {\bibfnamefont {L.}~\bibnamefont
  {Memarzadeh}}, \ and\ \bibinfo {author} {\bibfnamefont {P.}~\bibnamefont
  {Zarkeshian}},\ }\href {\doibase 10.1103/PhysRevA.87.032322} {\bibfield
  {journal} {\bibinfo  {journal} {Phys. Rev. A}\ }\textbf {\bibinfo {volume}
  {87}},\ \bibinfo {pages} {032322} (\bibinfo {year} {2013})}\BibitemShut
  {NoStop}%
\bibitem [{\citenamefont {Langari}\ \emph {et~al.}(2015)\citenamefont
  {Langari}, \citenamefont {Mohammad-Aghaei},\ and\ \citenamefont
  {Haghshenas}}]{Langari:2015}%
  \BibitemOpen
  \bibfield  {author} {\bibinfo {author} {\bibfnamefont {A.}~\bibnamefont
  {Langari}}, \bibinfo {author} {\bibfnamefont {A.}~\bibnamefont
  {Mohammad-Aghaei}}, \ and\ \bibinfo {author} {\bibfnamefont {R.}~\bibnamefont
  {Haghshenas}},\ }\href {\doibase 10.1103/PhysRevB.91.024415} {\bibfield
  {journal} {\bibinfo  {journal} {Phys. Rev. B}\ }\textbf {\bibinfo {volume}
  {91}},\ \bibinfo {pages} {024415} (\bibinfo {year} {2015})}\BibitemShut
  {NoStop}%
\bibitem [{\citenamefont {Bergman}\ \emph {et~al.}(2007)\citenamefont
  {Bergman}, \citenamefont {Shindou}, \citenamefont {Fiete},\ and\
  \citenamefont {Balents}}]{Bergman:2007}%
  \BibitemOpen
  \bibfield  {author} {\bibinfo {author} {\bibfnamefont {D.~L.}\ \bibnamefont
  {Bergman}}, \bibinfo {author} {\bibfnamefont {R.}~\bibnamefont {Shindou}},
  \bibinfo {author} {\bibfnamefont {G.~A.}\ \bibnamefont {Fiete}}, \ and\
  \bibinfo {author} {\bibfnamefont {L.}~\bibnamefont {Balents}},\ }\href
  {\doibase 10.1103/PhysRevB.75.094403} {\bibfield  {journal} {\bibinfo
  {journal} {Phys. Rev. B}\ }\textbf {\bibinfo {volume} {75}},\ \bibinfo
  {pages} {094403} (\bibinfo {year} {2007})}\BibitemShut {NoStop}%
\bibitem [{\citenamefont {Son}\ \emph {et~al.}(2011)\citenamefont {Son},
  \citenamefont {Amico}, \citenamefont {Fazio}, \citenamefont {Hamma},
  \citenamefont {Pascazio},\ and\ \citenamefont {Vedral}}]{son:2011}%
  \BibitemOpen
  \bibfield  {author} {\bibinfo {author} {\bibfnamefont {W.}~\bibnamefont
  {Son}}, \bibinfo {author} {\bibfnamefont {L.}~\bibnamefont {Amico}}, \bibinfo
  {author} {\bibfnamefont {R.}~\bibnamefont {Fazio}}, \bibinfo {author}
  {\bibfnamefont {A.}~\bibnamefont {Hamma}}, \bibinfo {author} {\bibfnamefont
  {S.}~\bibnamefont {Pascazio}}, \ and\ \bibinfo {author} {\bibfnamefont
  {V.}~\bibnamefont {Vedral}},\ }\href
  {http://iopscience.iop.org/0295-5075/95/5/50001/} {\bibfield  {journal}
  {\bibinfo  {journal} {Europhys. Lett.}\ }\textbf {\bibinfo {volume} {95}},\
  \bibinfo {pages} {50001} (\bibinfo {year} {2011})}\BibitemShut {NoStop}%
\bibitem [{\citenamefont {Else}\ \emph {et~al.}(2012)\citenamefont {Else},
  \citenamefont {Schwarz}, \citenamefont {Bartlett},\ and\ \citenamefont
  {Doherty}}]{Else:2012}%
  \BibitemOpen
  \bibfield  {author} {\bibinfo {author} {\bibfnamefont {D.~V.}\ \bibnamefont
  {Else}}, \bibinfo {author} {\bibfnamefont {I.}~\bibnamefont {Schwarz}},
  \bibinfo {author} {\bibfnamefont {S.~D.}\ \bibnamefont {Bartlett}}, \ and\
  \bibinfo {author} {\bibfnamefont {A.~C.}\ \bibnamefont {Doherty}},\ }\href
  {\doibase 10.1103/PhysRevLett.108.240505} {\bibfield  {journal} {\bibinfo
  {journal} {Phys. Rev. Lett.}\ }\textbf {\bibinfo {volume} {108}},\ \bibinfo
  {pages} {240505} (\bibinfo {year} {2012})}\BibitemShut {NoStop}%
\bibitem [{\citenamefont {Montes}\ and\ \citenamefont
  {Hamma}(2012)}]{Montes:2012}%
  \BibitemOpen
  \bibfield  {author} {\bibinfo {author} {\bibfnamefont {S.}~\bibnamefont
  {Montes}}\ and\ \bibinfo {author} {\bibfnamefont {A.}~\bibnamefont {Hamma}},\
  }\href {\doibase 10.1103/PhysRevE.86.021101} {\bibfield  {journal} {\bibinfo
  {journal} {Phys. Rev. E}\ }\textbf {\bibinfo {volume} {86}},\ \bibinfo
  {pages} {021101} (\bibinfo {year} {2012})}\BibitemShut {NoStop}%
\bibitem [{\citenamefont {Nussinov}\ and\ \citenamefont {van~den
  Brink}(2015)}]{Nussinov:2015}%
  \BibitemOpen
  \bibfield  {author} {\bibinfo {author} {\bibfnamefont {Z.}~\bibnamefont
  {Nussinov}}\ and\ \bibinfo {author} {\bibfnamefont {J.}~\bibnamefont {van~den
  Brink}},\ }\href {\doibase 10.1103/RevModPhys.87.1} {\bibfield  {journal}
  {\bibinfo  {journal} {Rev. Mod. Phys.}\ }\textbf {\bibinfo {volume} {87}},\
  \bibinfo {pages} {1} (\bibinfo {year} {2015})}\BibitemShut {NoStop}%
\bibitem [{\citenamefont {Wang}\ and\ \citenamefont {Cho}(2015)}]{Wang:2015}%
  \BibitemOpen
  \bibfield  {author} {\bibinfo {author} {\bibfnamefont {H.~T.}\ \bibnamefont
  {Wang}}\ and\ \bibinfo {author} {\bibfnamefont {S.~Y.}\ \bibnamefont {Cho}},\
  }\href {http://iopscience.iop.org/0953-8984/27/1/015603} {\bibfield
  {journal} {\bibinfo  {journal} {J. Phys.: Condens. Matter}\ }\textbf
  {\bibinfo {volume} {27}},\ \bibinfo {pages} {015603} (\bibinfo {year}
  {2015})}\BibitemShut {NoStop}%
\bibitem [{\citenamefont {Kargarian}\ \emph {et~al.}(2010)\citenamefont
  {Kargarian}, \citenamefont {Bombin},\ and\ \citenamefont
  {Martin-Delgado}}]{kargarian:2010}%
  \BibitemOpen
  \bibfield  {author} {\bibinfo {author} {\bibfnamefont {M.}~\bibnamefont
  {Kargarian}}, \bibinfo {author} {\bibfnamefont {H.}~\bibnamefont {Bombin}}, \
  and\ \bibinfo {author} {\bibfnamefont {M.}~\bibnamefont {Martin-Delgado}},\
  }\href {http://iopscience.iop.org/1367-2630/12/2/025018} {\bibfield
  {journal} {\bibinfo  {journal} {New J. Phys.}\ }\textbf {\bibinfo {volume}
  {12}},\ \bibinfo {pages} {025018} (\bibinfo {year} {2010})}\BibitemShut
  {NoStop}%
\bibitem [{\citenamefont {Hastings}(2007)}]{Hastings:2007}%
  \BibitemOpen
  \bibfield  {author} {\bibinfo {author} {\bibfnamefont {M.~B.}\ \bibnamefont
  {Hastings}},\ }\href {http://iopscience.iop.org/1742-5468/2007/08/P08024}
  {\bibfield  {journal} {\bibinfo  {journal} {J. Stat. Mech.: Theo. Exp.}\ ,\
  \bibinfo {pages} {P08024}} (\bibinfo {year} {2007})}\BibitemShut {NoStop}%
\bibitem [{\citenamefont {Or\'us}\ and\ \citenamefont
  {Vidal}(2008)}]{Orus:2008}%
  \BibitemOpen
  \bibfield  {author} {\bibinfo {author} {\bibfnamefont {R.}~\bibnamefont
  {Or\'us}}\ and\ \bibinfo {author} {\bibfnamefont {G.}~\bibnamefont {Vidal}},\
  }\href {\doibase 10.1103/PhysRevB.78.155117} {\bibfield  {journal} {\bibinfo
  {journal} {Phys. Rev. B}\ }\textbf {\bibinfo {volume} {78}},\ \bibinfo
  {pages} {155117} (\bibinfo {year} {2008})}\BibitemShut {NoStop}%
\bibitem [{\citenamefont {Haghshenas}\ \emph {et~al.}(2014)\citenamefont
  {Haghshenas}, \citenamefont {Langari},\ and\ \citenamefont
  {Rezakhani}}]{Haghshenas:2014}%
  \BibitemOpen
  \bibfield  {author} {\bibinfo {author} {\bibfnamefont {R.}~\bibnamefont
  {Haghshenas}}, \bibinfo {author} {\bibfnamefont {A.}~\bibnamefont {Langari}},
  \ and\ \bibinfo {author} {\bibfnamefont {A.~T.}\ \bibnamefont {Rezakhani}},\
  }\href {\doibase doi:10.1088/0953-8984/26/45/456001} {\bibfield  {journal}
  {\bibinfo  {journal} {J. Phys.: Condens. Matter}\ }\textbf {\bibinfo {volume}
  {26}},\ \bibinfo {pages} {456001} (\bibinfo {year} {2014})}\BibitemShut
  {NoStop}%
\bibitem [{\citenamefont {Perez-Garcia}\ \emph {et~al.}(2007)\citenamefont
  {Perez-Garcia}, \citenamefont {Verstraete}, \citenamefont {Wolf},\ and\
  \citenamefont {Cirac}}]{Garcia:2007}%
  \BibitemOpen
  \bibfield  {author} {\bibinfo {author} {\bibfnamefont {D.}~\bibnamefont
  {Perez-Garcia}}, \bibinfo {author} {\bibfnamefont {F.}~\bibnamefont
  {Verstraete}}, \bibinfo {author} {\bibfnamefont {M.~M.}\ \bibnamefont
  {Wolf}}, \ and\ \bibinfo {author} {\bibfnamefont {J.~I.}\ \bibnamefont
  {Cirac}},\ }\href {http://dl.acm.org/citation.cfm?id=2011832.2011833}
  {\bibfield  {journal} {\bibinfo  {journal} {Quantum Inf. Comput.}\ }\textbf
  {\bibinfo {volume} {7}},\ \bibinfo {pages} {401} (\bibinfo {year}
  {2007})}\BibitemShut {NoStop}%
\end{thebibliography}%
 
\end{document}